\documentclass[twocolumn,amsmath,amssymb,nofootinbib]{revtex4}

\usepackage{graphicx}
\usepackage{amssymb}
\usepackage{url}
\usepackage{physics}
\usepackage{placeins,color}

\newcommand{\me}[0]{\mathrm{e}}
\newcommand{\unit}[1]{\;\mathrm{#1}}
\newcommand{\iu}{\mathrm{i}}

\usepackage[pdftex,
       colorlinks=true,
       urlcolor=blue,       
        bookmarksopen=true]{hyperref}

\renewcommand{\vec}[1]{\boldsymbol{#1}}
\begin{document}

\title{The effect of non-equal emission times and space-time correlations on
  (anti-) nuclei production}

\author{M.~Kachelrie\ss$^1$}
\author{S.~Ostapchenko$^{2}$}
\author{J.~Tjemsland$^1$}

\affiliation{$^1$Institutt for fysikk, NTNU, Trondheim, Norway}
\affiliation{$^2$II. Institute for Theoretical Physics, Hamburg University, Hamburg,
  Germany}

\date{15.\ March 2023}

\begin{abstract}
Light (anti-) nuclei are a powerful tool both in collider physics
and astrophysics. In searches for new and exotic physics, the expected small
astrophysical backgrounds at low energies make these antinuclei ideal
probes for, e.g., dark matter. At the same time, their composite structure and
small binding energies imply that they can be used in collider experiments
to probe the hadronisation process and two-particle correlations. For the
proper interpretation of such experimental studies, an improved theoretical
understanding of (anti-) nuclei production in specific kinematic regions and
detector setups is needed.
In this work, we develop a coalescence framework for (anti-) deuteron
production which accounts for both the emission volume and momentum
correlations on an event-by-event basis: While momentum correlations can
be provided by event generators, such as PYTHIA, the emission volume
has to be  derived from semi-classical considerations. Moreover, this
framework goes beyond the equal-time approximation, which has been often
assumed in femtoscopy experiments and (anti-) nucleus production models
until now in small interacting systems. 
Using PYTHIA~8 as an event generator, we find that the equal-time approximation
leads to an error of $\mathcal{O}(10\%)$ in low-energy processes like
$\Upsilon$ decays, while the errors are negligible at LHC energies. The
framework introduced in this work paves the
way for tuning event generators to (anti-) nuclei measurements.
\end{abstract}


\maketitle

\section{Introduction}

Light (anti-) nuclei are interesting particles due to their composite
structure and small binding energies. This makes them ideal probes for,
e.g., two-particle correlations and the QCD phase diagram in heavy ion
collisions~\cite{Caines:2017vvs}. In particle collisions and decays,
(anti-) nuclei can provide valuable information on
the hadronisation process and momentum correlations that can be used to
tune QCD inspired event generators. For the astroparticle community,
the production of antinuclei is of immense interest since it is an ideal
tool to search for new and exotic physics, such as dark matter annihilations
or decays in
the Milky Way~\cite{Chardonnet:1997dv,Donato:1999gy,vonDoetinchem:2020vbj}.
In order to correctly interpret astrophysical and collider data, a description
of the formation process as precise as possible is desirable.

The best motivated production model for light nuclei\footnote{In the following,
  we denote with nuclei both nuclei and anti-nuclei.}  in particle
collisions---especially for small interacting systems---is arguably the
coalescence model. In this model, final-state nucleons may merge if they are
close in phase space. In heavy-ion collisions, the coalescence probability
is often assumed to be mainly determined by the nucleon emission volume,
while momentum correlations are neglected or treated as a collective
effect~\cite{Csernai:1986qf,Nagle:1996vp}. In small interacting systems, on
the other hand, the coalescence condition is typically only evaluated in
momentum space: For instance, in the simplest phenomenological coalescence
model two nucleons merge if the momentum difference $\Delta p$ in their
pair rest frame is smaller than the coalescence momentum
$p_0$~\cite{Schwarzschild:1963zz,butler_deuterons_1963}.
However, two-particle correlations should not be neglected in small systems
because of the low multiplicities and large anti-correlations of
produced nucleons~\cite{Gustafson:1993mm}. It was therefore suggested in
Refs.~\cite{Dal_thesis,kadastik_enhanced_2010} that
the coalescence condition $\Delta p<p_0$ should be evaluated on an
event-by-event basis using a Monte Carlo event generator.
Moreover, the expected nucleon emission length in small interacting
systems, $\sigma\simeq 1\unit{fm}$, is of the same order as the
size of the  wave function of the deuteron,
 ${r_\mathrm{rms}^d\simeq 2\unit{fm}}$,
even in point-like interactions~\cite{Kachelriess:2019taq}~(see also
Ref.~\cite{Gustafson:1993mm} for an early discussion of the decay of
$\Upsilon$ mesons). Thus, one should consider both the size of the formation
region and momentum correlations on an event-by-event basis
 simultaneously.
This is currently only achieved by the WiFunC model (Wigner Function with
Correlations) introduced in Ref.~\cite{Kachelriess:2019taq}, and further
developed and discussed in Refs.~\cite{Kachelriess:2020uoh,%
Kachelriess:2020amp,Tjemsland:2020bzu}. This model is especially suitable
for production processes relevant to cosmic ray
interactions~\cite{Kachelriess:2020uoh,Kachelriess:2022khq,Serksnyte:2022onw}.

The WiFunC model, as most other sophisticated coalescence models~\cite{scheibl_coalescence_1999,Sun:2018jhg,Sun:2017xrx,Shao:2019xpj,Sun:2020uoj,Zhao:2021dka,Shao:2022eyd}, 
relies on the Wigner function approach, in which the coalescence
probability is found by projecting the nucleon Wigner function onto the
Wigner function of the light
nucleus~\cite{Gyulassy:1982pe,Nagle:1996vp,Danielewicz:1991dh,Mattiello:1996gq}.
One of the key advantages of this approach is the fact that the
coalescence probability depends on the hadronic emission region, a quantity
which can be measured in femtoscopy experiments~\cite{scheibl_coalescence_1999}.
This allows one to determine independently the free parameter of these
models~\cite{Blum:2017qnn,Blum:2019suo,Kachelriess:2020amp}.
Moreover, femtoscopy experiments can be used to distinguish between the
coalescence hypothesis and other formation processes like thermal
freeze-out~\cite{Acharya:2017bso,Andronic:2017pug,Vovchenko:2018fiy,%
  Bellini:2018epz,Chen:2018tnh,Xu:2018jff,Oliinychenko:2018ugs}.
For instance, it was argued in Ref.~\cite{Bellini:2020cbj} that the current
success of the framework is a strong indication that coalescence is a major
antinuclei production mechanism.

This work is structured as follows:
In section~\ref{sec:theory}, we review the basis of the Wigner function
approach to coalescence, focusing on small interacting systems. In particular,
we extend the framework to allow for non-equal emission times of the nucleons.
That is, we go beyond the equal-time approximation which underlies both the
experimental and theoretical framework of femtoscopy, and is expected to give
up to $\mathcal{O}(30\unit{\%})$ uncertainties~\cite{Bellini:2020cbj}.
This effect has previously been considered
in the context of heavy-ion collisions when momentum correlations can
be neglected, in which case the source radius effectively is increased
as $r\to r+vt$~\cite{Mrowczynski:1992gc,Lednicky:1995vk,Maj:2009ue},
or using transport codes~\cite{Nagle:1994hm,Mattiello:1996gq}.
In this work, however, we are interested in small interacting systems
where momentum correlations should not be neglected.
In section~\ref{sec:wifunc}, we review the WiFunC model and give an in-depth
discussion of the choice of the nucleus wave function.
Furthermore, we discuss the possibility of using the semi-classical space-time
picture in QCD inspired event generators to describe the nucleon emission
volume, thus allowing one to take into account space-time correlations on
an event-by-event basis.
Finally, in
section~\ref{sec:examples}, the discussions are exemplified using Pythia
8.3~\cite{Bierlich:2022pfr,Ferreres-Sole:2018vgo}, with a focus on the
equal-time approximation and the
space-time picture provided by Pythia. Concretely, we consider
the size of the hadronic emission region~\cite{Acharya:2020dfb} (in~\ref{sec:volume}),
the antideuteron spectrum~\cite{ALICE:2017xrp} (in~\ref{sec:spectrum}), 
and the coalescence probability in
jets~\cite{ALICE:2022jmr,ALICE:2020hjy} (in~\ref{sec:jets})
measured by the ALICE collaboration.
Furthermore, we compute the energy dependence of the emission volume,
predicted by Pythia in section~\ref{sec:energy}
and the antideuteron production
in $\Upsilon$ decays~\cite{BaBar:2014ssg,CLEO:2006zjy} in
section~\ref{sec:upsilon}.
The examples indicate that the equal-time approximation leads to an error of
$\sim10\%$ at low energies, while the error is neglible at LHC energies. 
The main
uncertainties of the WiFunC model and of its predictions are currently related to
the accuracy of the underlying event generators for high energy collisions.
Conversely, the framework allows one to use
femtoscopy and antideuteron measurements to tune such event generators.

\section{The Wigner function approach to coalescence}
\label{sec:theory}

\subsection{General frame-work}

In femtoscopy experiments, the correlations of pairs of particles with small
relative momenta are measured. Since the final-state interactions that give
rise to the correlations even from an initially uncorrelated source decrease
rapidly with increasing  distance in phase space, we only consider the
contribution from the dominant pair\footnote{This was checked explicitly in
Ref.~\cite{Kachelriess:2019taq} for coalescing nucleons in small interacting
systems.}. The double energy spectrum can in this case be written as 
\begin{widetext}
\begin{equation}
  (2\pi)^8\gamma_1 \frac{\dd[6]N}{\dd[3]p_1\dd[3]{p_2}} =
    \sum_{S}\int \dd[4]{x_1} \dd[4]{x_2}\dd[4]{x_1'} \dd[4]{x_2'}
    \rho(x_1, x_2; x_1', x_2')
    \Psi^{S(-)}_{d, P}(x_1, x_2) {\Psi^{S(-)}_{d, P}}^{\dagger}(x_1, x_2) ,
  \label{eq:lednicky}
\end{equation}
where $\rho$ is the two-particle density matrix of the source and
$\Psi_{d, P_d}^{S(-)}(x_1,x_2) = [\Psi_{d, P_d}^{S(+)}(x_1, x_2)]^\dagger$ is the
Bethe-Salpeter wave function accounting for the final-state
interactions~\cite{Lednicky:2005tb}.
In the case of weakly bound systems such as the  deuteron, 
helion and triton, we can connect Eq.~\eqref{eq:lednicky} with the
coalesence formalism based on generalised or relativistic Wigner functions:
Neglecting the binding energy and employing the sudden approximation, the
Bethe-Salpeter wave function reduces to the wave function of the static bound
state. The  deuteron energy spectrum can then be approximated as
\begin{equation}
  (2\pi)^8\gamma_d \dv[3]{N_d}{P_d} =
    S\int \dd[4]{x_1} \dd[4]{x_2}\dd[4]{x_1'} \dd[4]{x_2'}
    \rho(x_1, x_2; x_1', x_2')
    \Psi^{(-)}_{d, P_d}(x_1, x_2) {\Psi^{(-)}_{d, P_d}}^{\dagger}(x_1, x_2),
  \label{eq:init}
\end{equation}
where the factor $S=3/8$ is obtained by averaging over all spin and isospin
states. Note that this formula follows directly from the general rules of
(relativistic) statistical quantum mechanics, if the sudden approximation
is employed. The latter requires that the formation time $\tau$
of the deuteron can be neglected relative to the inverse of its binding energy
$E_d$~\cite{Schiff63}, i.e.\ that $\tau\ll 1/E_d\simeq 90$\,fm.
Factoring out then the center-of-mass motion,
$\Psi_{d, P_d}^{S(-)}(x_1,x_2)=\me^{\iu P_d X}\Psi_d^{S(-)}(x)$,
one can re-write Eq.~\eqref{eq:init}) as
(see, e.g., Ref.~\cite{scheibl_coalescence_1999} for details),
\begin{equation}
  (2\pi)^8\gamma_d \dv[3]{N_d}{P_d} =
    S\int\dd[4]{r}\dd[4]{r_d}\dd[4]{q}\mathcal{D}^{(4)}(q,r)
    W_{np}(P_d/2+q,P_d/2-q,r,r_d),
  \label{eq:wifunc}
\end{equation}
where
\begin{equation}
  \mathcal{D}^{(4)}(q, r) = \int \dd[4]{\xi}\me^{-\iu q\xi}
        \Psi_d^{(-)}(r+\xi/2){\Psi_d^{(-)}}^*(r-\xi/2)
  \label{eq:deuteron_wigner_function}
\end{equation}
is the (generalised or off-shell) deuteron Wigner function and $W_{np}$ the
two-nucleon Wigner function of the source.
Here, $r$ denotes the space-time distance between the nucleons, $r_d$ the
space-time position and $P_d=p_1+p_2$  the
four-momentum of the nucleus, while $q=(p_1-p_2)/2$ is  the four-momentum of
the nucleons in the nucleus frame. The main difference between
Eq.~\eqref{eq:wifunc} and the expression usually used in the literature (e.g.\
in Refs.~\cite{scheibl_coalescence_1999,Kachelriess:2019taq,Bellini:2020cbj})
is its time and energy dependence: The variable
$r_d^0$ describes the ``freeze-out'' time of the nucleus, and does not affect
the emission volume\footnote{We emphasise that we are referring to
the emission volume as a function of $r$, which is what one measures
experimentally.}.
Meanwhile, the variable $r^0=t$ describes the
time difference in the production of the nucleons, which clearly
may impact the measured emission volume. Finally, $q^0$ describes the
off-shell structure of the two-particle system.

To proceed, it is normally assumed that the particles are produced at the
same time (equal-time approximation) and/or that the source is independent of
$q$ (smoothness-approximation)~\cite{Lisa:2005dd}.
However, as argued in Ref.~\cite{Bellini:2020cbj},
the equal-time approximation is not expected to be accurate
in the case of small interacting systems. In order to
check the reliability of this approximation, one should therefore go beyond
the equal time approximation. As we will see, it is sufficient to assume
that the particles are (approximately) on-shell when they coalesce, 
$W_{np}\simeq W_{np}(q^0=0)$. This assumption is well motivated due to the
low binding energy of the antinuclei. 
We are in this case left with\footnote{We define
  $W^{(4)}(\vec{P}_d/2+\vec{q},\vec{P}_d/2-\vec{q},r)=
  \int \dd[4]{r_d} W_{np}^{(4)}
  (\vec{P}_d/2+\vec{q},\vec{P}_d/2-\vec{q},r,r_d)$
}
\begin{equation}
    (2\pi)^7\gamma_d \dv[3]{N_d}{P_d} =
    S\int\dd[4]{r}\dd[3]{q}\mathcal{D}^{(3)}_t(\vec{q},\vec{r})
    W_{np}^{(4)}(\vec{P}_d/2+\vec{q},\vec{P}_d/2-\vec{q},r),
    \label{eq:main0}
\end{equation}
where
\begin{equation}
  \mathcal{D}^{(3)}_t(\vec{q},\vec{r})=\int \dd[3]{\xi}
    \me^{-\iu \vec{q}\cdot\vec{\xi}}
      \Psi_d^{(-)}(\vec{r}+\vec{\xi}/2, t)
      {\Psi_d^{(-)}}^*(\vec{r}-\vec{\xi}/2,t)
  \label{eq:wig2}
\end{equation}
\end{widetext}
is the time dependence of the static deuteron Wigner function. In the
nucleus frame, this reduces to
\begin{equation}
  \dv[3]{N_d}{P_d} =
    \frac{S}{(2\pi)^7}\int\dd{t}\dd[3]{r}\dd[3]{q}
    \mathcal{D}^{(3)}_t(\vec{q},\vec{r})
    W_{np}^{(4)}(\vec{q},\vec{r}, t).
\end{equation}
In order to evaluate the deuteron yield using Eq.~\eqref{eq:main0}, the deuteron
Wigner function and the two-nucleon Wigner function have to be modelled. A key
observation is that, in the classical limit, the nucleon Wigner function will
reduce to  the phase-space distribution.
In section~\ref{sec:wifunc}, we discuss the WiFunC approach, in which the
momentum correlations are provided by an event generator, while the emission
volume is either assumed to be Gaussian or taken also 
from an event generator.

 \subsection{The effect of non-equal emission times}

The various coalesence models based on the  Wigner function approach
differ mainly in the way how these functions are determined:
In heavy-ion collisions, semi-classical transport models like
the RQMD~\cite{Sorge:1989vt} or AMPT~\cite{Sun:2020uoj} schemes are
used to describe the
space-time evolution of the
particles~\cite{Nagle:1994hm,Mattiello:1996gq}. 
While quantum effects are
included via Pauli blocking and the stochastic nature of scatterings,
the propagation of particles proceeds in these schemes classically.
In contrast, many approaches which aim to describe coalescence and
femtoscopy experiments in smaller interacting systems prefer
to stay as long as possible within the realm of quantum mechanics. Therefore
they have to rely typically  on the equal-time
approximation~\cite{Lisa:2005dd,Bellini:2020cbj}, i.e., they assume
that the particles are produced at the same time, $t=0$.
More concretely, it is assumed that
$q\ll m \sigma/t\simeq 1$\,GeV~\cite{Lednicky:2005tb},
where $\sigma$ is the linear size of the emission volume.
Since the bulk of nuclei are produced by nucleons with
$q\sim \mathcal{O}(0.1)$\,GeV, this condition is expected to yield
an uncertainty of $\mathcal{O}(10\%)$
in $pp$ collisions~\cite{Bellini:2020cbj}. 

The effect of non-equal emission times on femtoscopy experiments is discussed
in detail in Ref.~\cite{Lednicky:2005tb}, where it is shown that the relation
between the Bethe-Salpeter amplitude and the corresponding non-relativistic wave
function can be expressed as
\begin{equation}
  \Psi(r)=\Psi(\vec{r}, t)=\int \dd[3]{r'} \delta_q(\vec{r}-\vec{r}', t)
          \psi(\vec{r}')
  \label{eq:lednicky_relation}
\end{equation}
under the condition $q^2\ll m^2$, which clearly is the case we are interested
in. The function $\delta_q(\vec{r}, t)$ reduces to the ordinary Dirac
delta function for $t=0$, 
and for $t>0$ it is given by~\cite{Lednicky:2005tb}
\begin{equation}
  \delta_q(\vec r, t) = \left(\frac{m}{2\pi \iu t}\right)^{3/2}
    \exp\left(\frac{\iu q^2 t}{2m} + \frac{\iu \vec r^2 m}{2t}\right).
  \label{eq:delta}
\end{equation}
Inserting Eq.~\eqref{eq:lednicky_relation} into Eq.~\eqref{eq:wig2} leads to
\begin{equation}
  \mathcal{D}_t^{(3)}(\vec q, \vec r) =
  \mathcal{D}^{(3)}(\vec q, \vec r+\vec qt/m)
\end{equation}  
with
\begin{equation}
    \mathcal{D}^{(3)}(\vec q, \vec r) = \int\dd[3]\xi
        \me^{-\iu\vec{\xi}\cdot \vec q}
        \psi(\vec r + \vec\xi/2)\psi^*(\vec r - \vec{\xi}/2).
  \label{eq:deut_wigner}
\end{equation}
Therefore, the deuteron yield can be expressed as
\begin{equation}
  \dv[3]{N_d}{P_d} =
    \frac{S}{(2\pi)^7}\int\dd{t}\dd[3]{r}\dd[3]{q}
    \mathcal{D}^{(3)}(\vec{q},\vec{r}+\vec qt/m)
    W_{np}^{(4)}(\vec{q},\vec{r}, t)
    \label{eq:main}
\end{equation}
in the pair rest frame. 
By comparing with, e.g.,
Refs.~\cite{scheibl_coalescence_1999,Kachelriess:2019taq,Bellini:2020cbj}, one
can see that non-equal emission times of the nucleons 
change $r$ by the classical distance the first particle propagates before
the second particle is produced.
If the equal-time approximation ($t\to 0$) is applied to Eq.~\eqref{eq:main},
one re-obtains, as expected, the same equation as
in Refs.~\cite{scheibl_coalescence_1999,Kachelriess:2019taq,Bellini:2020cbj}.

Note that  four assumptions are needed to obtain
Eq.~\eqref{eq:main}: (1) The coalescing particles are non-relativistic in the
pair rest frame ($q^2\ll m^2$) and (2) approximately on-shell. Moreover, (3)
the wave function describing the initial and final states changes slowly
compared to the interaction time (i.e., the sudden approximation) and  (4)
the interaction between a single pair of nucleons is dominant. All these
assumptions are well motivated, and always used
in the coalescence model. For example, due to the
small binding energy of the deuteron, one will expect that the nucleons have
to be close in phase space and approximately on-shell to coalesce.

\subsection{Relation to femtoscopy experiments}
\label{sec:femtoscopy}

Since the measured source function is strongly linked to the Wigner function,
any coalescence model arising from Eq.~\eqref{eq:main0} can be directly and
independently tested and tuned by baryonic correlation experiments.
Under the smoothness approximation (i.e.
$W_{np}(r, r_d, P_d, q)\simeq W_{np}(r, r_d, P_d, 0)$), Eq.~\eqref{eq:main0}
can be written as
\begin{align}
  \dv[3]{N_d}{P_d} & \propto \int \dd[3]{r} \dd{t} 
    W_{np}^{(4)}(\vec{r}, t) \int \dd[3]{q}
    \mathcal{D}_t^{(3)}(\vec q, \vec r)
\nonumber\\ &  = \int \dd[3]{r} |\phi(\vec{r})|^2 \mathcal{S}(\vec{r}),
  \label{eq:femtoscopy}
\end{align}
where $\mathcal{S}(\vec{r}) = W_{np}^{(3)}(\vec{r})$ is the emission source
defined in the pair rest frame. The last equality follows directly
if one in addition uses the equal-time approximation~\cite{Lisa:2005dd},
$W_{np}\propto \delta(t)$.
In a femtoscopy experiment, the source size $\mathcal{S}(\vec{r})$ can be
measured via the final-state interactions encoded into the wave function
$\psi(r)$~\cite{Lisa:2005dd}. Thus, a femtoscopy experiment can be interpreted
as an indirect measurement of the Wigner function.
Recently, the ALICE collaboration measured the size of the baryonic emission
source in  $pp$ collisions at
13\,TeV, assuming an isotropic Gaussian source~\cite{Acharya:2020dfb}. This
measurement can be used to fix the free parameter of a coalescence model,
allowing one to test and tune the coalescence models like the WiFunC
model~\cite{Kachelriess:2020amp}.

Femtoscopy experiments include often a cut in the
momentum $q$~\cite{Acharya:2020dfb}. It is thus interesting to 
note that it is sufficient to assume that
$qt/m\ll r$ to derive Eq.~\eqref{eq:femtoscopy}, thereby removing
the need to invoke the equal-time and smoothness approximations.

\section{The WiFunC model}
\label{sec:wifunc}

In the classical limit, the nucleon Wigner function [see Eq.~\eqref{eq:main}]
will describe the phase space distribution of the
nucleons~\cite{doi:10.1119/1.2957889}. 
The main idea behind the WiFunC model is to include particle momentum
correlations  provided by a Monte Carlo event generator. At the same time,
the nucleon emission volume can be described either by
an ansatz, following general arguments regarding time and distance scales
in the production process,
or by the event generator. 
In this section, we give a short review of the
model and at the same time  a deeper discussion of the choice of the
nucleus wave function as well as the use of the space-time picture provided by
an event generator to describe the nucleon emission volume. In particular, we
comment on the consequences of the equal-time approximation
[cf. Eq.~\eqref{eq:main}].

\subsection{Deuteron wave function}

\begin{figure}
  \centering
  \includegraphics[scale=0.5]{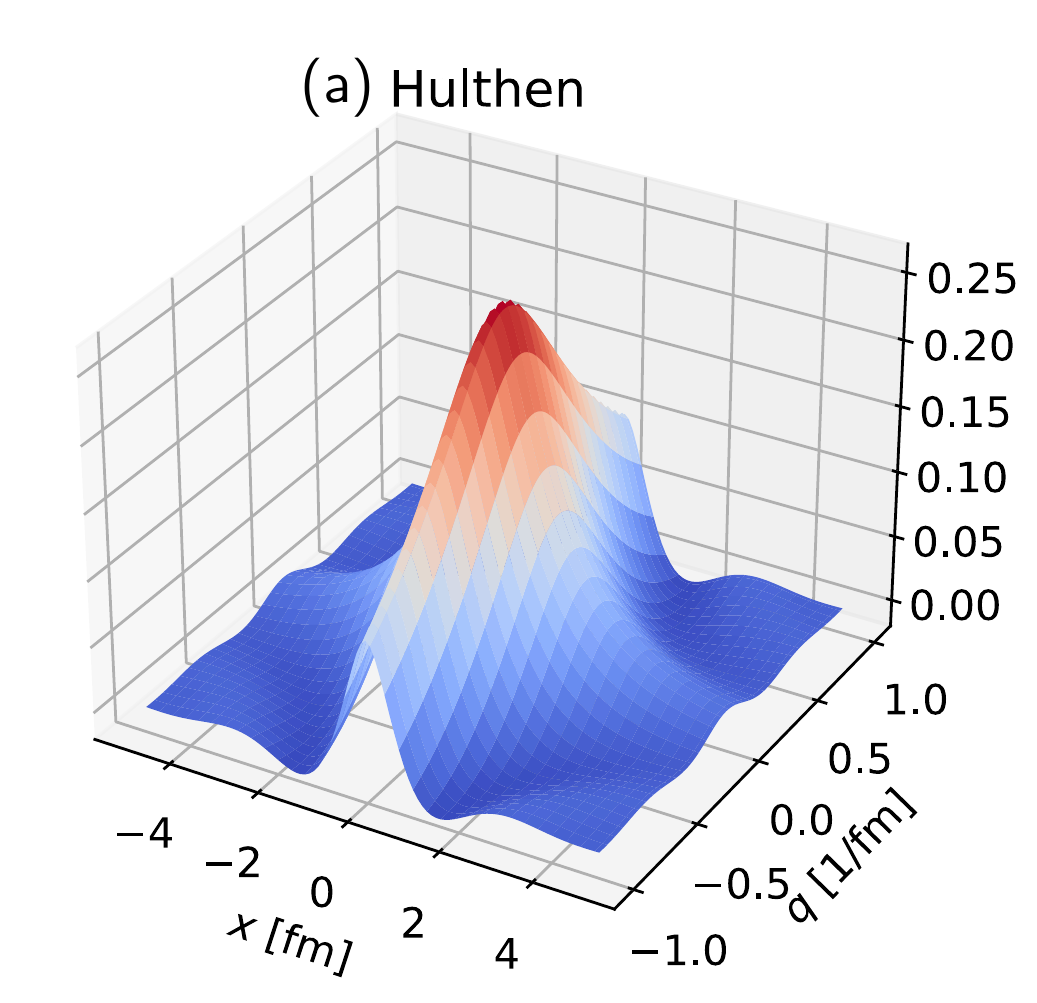}
  \includegraphics[scale=0.5]{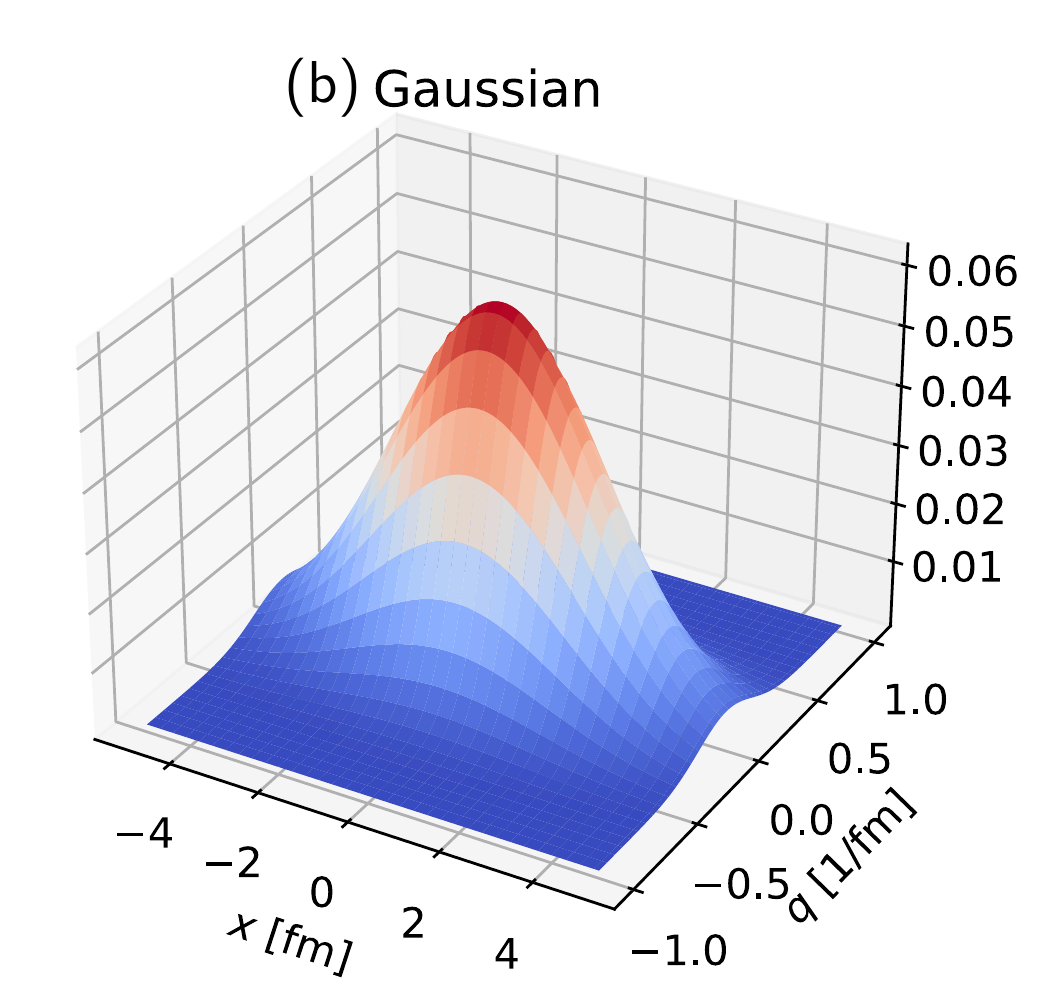}
  \includegraphics[scale=0.5]{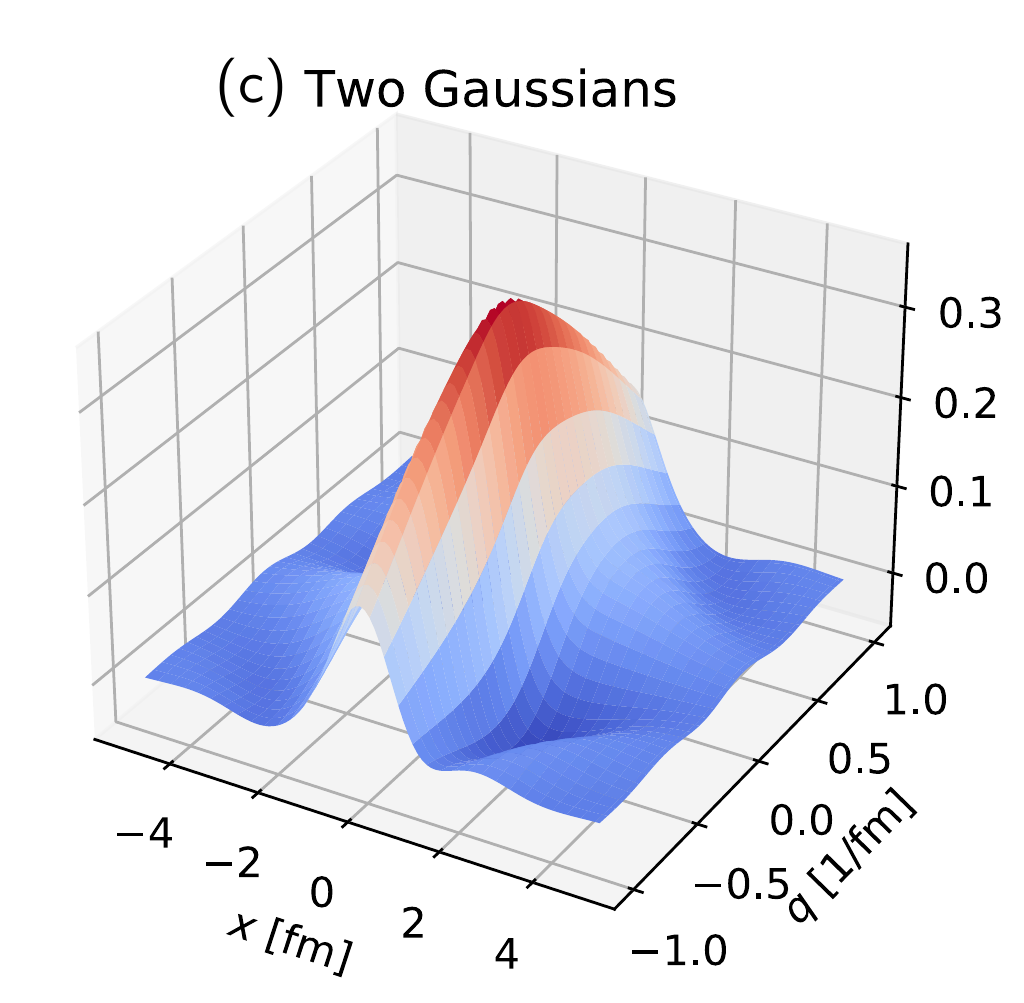}
  \includegraphics[scale=0.5]{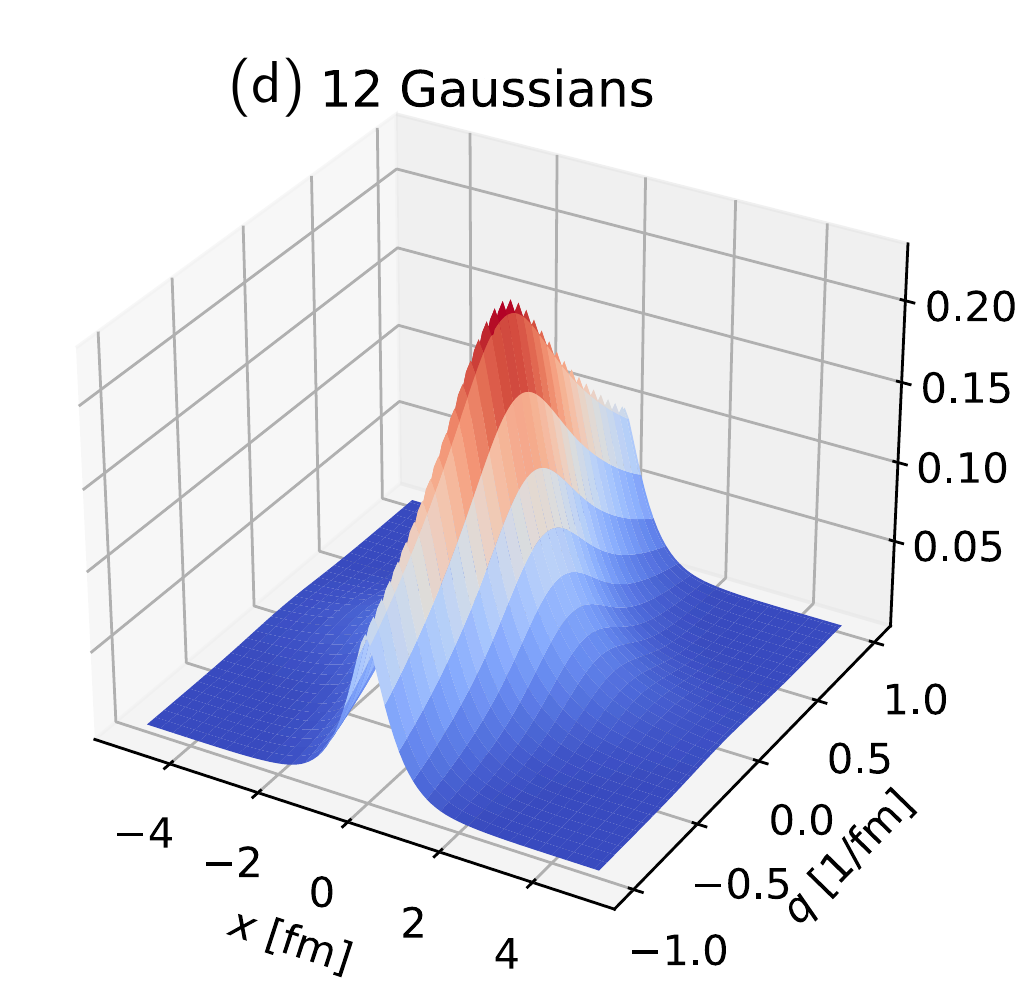}
  \caption{The Wigner function in the $(x,q_x)$ plane obtained numerically
   using (from top to bottom) (a) the Hulthen wave function,
   (b) a single Gaussian
   wave function, (c) the sum of two Gaussians with a phase shift $\pi$, and
   (d) the
   sum of 12 Gaussians. The wave functions are discussed in more detail in the
   text and in Ref.~\cite{Kachelriess:2019taq}. While it is difficult to see
   in the plot,
   the Hulthen Wigner function is symmetric, while the negative parts of the
   two-Gaussian Wigner function are antisymmetric.}
  \label{fig:wigner}
\end{figure}

In the WiFunC model, the deuteron Wigner function $\mathcal{D}$ is an
essential ingredient in the calculation of the coalescence probability.
For a specific choice of the deuteron wave function $\phi$, the corresponding
Wigner function $\mathcal{D}$ can be evaluated using Eq.~\eqref{eq:deut_wigner}.
The deuteron is in a pure state, and can be well approximated by the Hulthen
wave function~\cite{Zhaba:2017syr}.
However, it is
known that the Wigner function of a pure state is strictly positive if and
only if the wave function is a
Gaussian~\cite{HUDSON1974249,doi:10.1063/1.525607}. An
interpretation of the deuteron Wigner function as a probability distribution,
as it is required for the evaluation of the coalescence probability, requires 
therefore at first sight to use a simple Gaussian wave function,
$\phi(r)\propto \exp\left(-r^2/2d^2\right)$.
In this case, the Wigner function becomes
\begin{equation}
  \mathcal{D}(r, q) = 8\me^{-r^2/d^2 - d^2q^2}.
\end{equation}
where the choice $d=3.2\unit{fm}$ reproduces the deuteron charge radius.
However,
the Gaussian wave function is neither a good representation of the Hulthen wave
function nor does it lead to a Wigner function which
is similar to that obtained using the Hulthen wave function, cf.\ with
Fig.~\ref{fig:wigner}. Thus, one should aim for a better description
of the deuteron wave function.

In order to find such an improved wave function, consider now the more
general pure state $\phi(r) = (\phi_r(r) + \iu\phi_i(r))/\sqrt{2}$,
where $\phi_r$ and $\phi_i$ are real wave-functions. In this case, the Wigner
function can be split into a symmetric and an antisymmetric part,
$\mathcal{D}(q, r)= \mathcal{D}_r(q,r) + \mathcal{D}_i(q,r) - A(q, r)$,
where $\mathcal D_r$ and $\mathcal{D}_i$ are the Wigner functions of
$\phi_r$ and $\phi_i$, respectively. The antisymmetric interference
term $A$ vanishes upon performing the integrations in Eq.~\eqref{eq:main}
and will therefore not contribute to the coalescence probability. 
This implies that the sum of two Wigner functions from pure states can be
re-cast into a Wigner function from a mixed state,
\begin{align}
\lefteqn{  \mathcal{D}_r(\vec q,\vec r) + \mathcal{D}_i(\vec q,\vec r)  =
    \int\dd[3]{\xi}
  \me^{-\iu\vec{q}\cdot\vec{\xi}}}
  \nonumber\\ & \times
    \bra{\vec r - \vec{\xi}/2}
    \left(\frac{1}{2}\ket{\phi_r}\bra{\phi_r}
      +\frac{1}{2}\ket{\phi_i}\bra{\phi_i}\right)
    \ket{\vec r + \vec{\xi}/2} .
    \label{eq:sum_of_wigner_functions}
\end{align}
A particular choice is the ``$\phi_0$-fit'' of Ref.~\cite{Kachelriess:2019taq},
where the deuteron Wigner function is given by
\begin{equation}
  \mathcal{D}(\vec{q}, \vec{r}) = 8\Delta \me^{-d_1^2q^2-r^2d_1^2} +
      8(1-\Delta) \me^{-d_2^2q^2-r^2d_2^2} - \mathcal{A}(\vec q, \vec r)
      \label{eq:phi0}
\end{equation}
with $\Delta=0.581$, $d_1=3.979\,$fm, $d_2=0.890\,$fm and
$\mathcal{A}$ is antisymmetric in $\vec{q}$ and $\vec{r}$.

If one describes the deuteron---incorrectly---as a mixed state, one can
approximate its wave function, e.g., the Hulthen wave function, arbitrarily
accurately by a sum of Gaussian states. In the fourth panel of
Fig.~\ref{fig:wigner}, we show a one-dimensional example\footnote{
  See, e.g., Ref.~\cite{Mattiello:1996gq} for an example in 3D.
} using 12 Gaussians whose centers
are distributed evenly between $x=-5$ and $x=5$.
Since we are considering a mixed state of Gaussians, the deuteron Wigner
function is itself a sum of Gaussians and strictly positive. This
approach presents a clear method for handling the negative parts of the
phase space distribution. However, the mixed state neglects the 
``quantum correlations'' encoded in the Wigner function.

The negative parts of the Wigner function should vanish in the classical limit.
This leads to another method of getting around the problem of a
negative Wigner function: If one uses the
equal-time approximation and assumes that the space and momentum distributions
are uncorrelated, $W_{np}=G(\vec q)H(\vec r_p, \vec r_n)$, the deuteron
yield can be written as
\begin{align}
  \dv[3]{N_d}{P_d} & =
    \frac{S}{(2\pi)^6}\int\dd[3]{q}
    G_{np}^{(3)}(\vec{q})\int\dd[3]{r_p}\dd[3]{r_n}
    \nonumber\\ & \times
    \mathcal{D}^{(3)}(\vec{q},\vec{r}) H_{np}(\vec{r}_n, \vec{r}_p)
    \label{eq:p}
\end{align}
where the last integral can be interpreted at the probability density for
coalescence and $G_{np}$ is the momentum distribution provided by the event
generator~\cite{Kachelriess:2019taq}.
These are the same assumptions used in the next subsection, where $H_{np}$ is
approximated as a Gaussian.
If $H_{np}$ is sufficiently wide and
well-behaved, the ``probability density'' will be strictly positive\footnote{
  The positivity condition depends on the shape of $H_{np}$ and the wave
  function.
}. Thus, this may allow one to use \emph{any} wave function and evaluate
numerically the coalescence probability event-by-event.
While this may work well for, e.g., high
multiplicity $pp$ collisions, the method should not be
applied if position--momentum correlations are included, or the
multiplicity of the interaction is small, such as in $\Upsilon$ decays.

In conclusion, if the deuteron wave function in the WiFunC model is represented
by \emph{any} $\phi(\vec{r})=\phi_r(\vec{r})+\iu \phi_i(\vec{r})$,
where $\phi_r$ and $\phi_i$ are Gaussians centered at $r=0$, the
coalescence probability is well defined for all interactions; we suggest,
with the current theoretical uncertainties,
using the Wigner function in Eq.~\eqref{eq:phi0}.

\subsection{Nucleon distribution}
\label{sec:gaussian}

Current QCD inspired event generators evaluate the parton cascade
in momentum space, using a probabilistic scheme. While this is sufficient
to provide two-particle momentum correlations, an extraction of the
two-nucleon Wigner function $W_{np}$ is not possible. Therefore, a
semi-classical ansatz has to be made before one can
evaluate the coalescence equation \eqref{eq:main} on an event-by-event basis.
In Ref.~\cite{Kachelriess:2019taq}, the equal-time approximation was used and
it was assumed that the space and momentum
distributions are uncorrelated, $W_{np}(\vec q, \vec P_d, \vec r_p, \vec r_n)
=G(\vec q)H(\vec r_p, \vec r_n)$. In turn, the
ansatz
\begin{equation}
  H(\vec{r}_p, \vec{r}_n) = h(\vec{r_p})h(\vec{r_n})
\end{equation}
with
\begin{equation}
  h(\vec{r}) = \left(2\pi\sigma^2\right)^{-3/2}
  \exp\left(-\frac{r_\perp^2}{2\sigma_\perp^2}-
    \frac{r_\parallel^2}{2\sigma_\parallel^2}\right)
\end{equation}
was used for the nucleon distributions in the laboratory (lab) frame.
In particle collisions, e.g., $pp$, $e^+e^-$ and $pN$, the longitudinal and
transverse directions are defined relative to the beam direction. In 
annihilation and decay processes, e.g., dark matter annihilations, one should
define the coordinate system relative to the initial quark--antiquark pair.
With a one-Gaussian
wave function, the deuteron spectrum can be written as
\begin{equation}
  \dv[3]{N_d}{P_d}=\frac{3\zeta(d)}{(2\pi)^3}
  \int\dd[3]{q}\mathrm{} \mathrm{e}^{-q^2d^2} G(\vec q),
  \label{eq:wifunc0}
\end{equation}
where
\begin{equation}
  \zeta(d) = \left[\left(\frac{d^2}{d^2 + 4\sigma_\perp^2 m_T^2/m^2}\right)
   \!\! \left(\frac{d^2}{d^2+4\sigma_\perp^2}\right)
   \!\! \left(\frac{d^2}{d^2+4\sigma_\parallel^2}\right) \right]^{1/2} \!\! .
  \label{eq:zeta}
\end{equation}
The $m_T$ dependence arises due to the Lorentz boost of the transverse spread
from the lab frame to the pair rest frame, see Ref.~\cite{Kachelriess:2019taq}
for details.  
The model can be added as an afterburner to any Monte Carlo event generator by
applying the weight
\begin{equation}
  w = 3   \Delta \zeta(d_1)\mathrm{e}^{-d_1^2q^2} +
        3(1-\Delta)\zeta(d_2)\mathrm{e}^{-d_2^2q^2},
    \label{eq:weight_gaussian}
\end{equation}
to each nuleon pair. Here, the numerical values of the parameters are
$\Delta=0.581$, $d_1=3.979$\,fm and $d_2=0.890$\,fm, while
$m_T$ and $q$ are determined event-by-event from the Monte Carlo data.

The two parameters $\sigma_i$ describe the average emission length of nucleons,
$\sigma_{\parallel/\perp}\simeq 1$\,fm.
In point-like processes, like $e^+e^-$ collisions, the longitudinal spread is
dominated by the hadronisation length,
$\sigma_\parallel\sim L_\mathrm{had}\simeq 1$\,fm, while the transverse
spread is related to $\Lambda_\mathrm{QCD}$. Since they are of the same
order of magnitude, it is convenient to set
$\sigma=\sigma_\perp=\sigma_\parallel$. In collisions involving hadrons and
nuclei, the
spread will also obtain a geometrical contribution due to multiple
parton--parton
scatterings. In the particular case of $pp$ collisions, the spread in the
transverse and longitudinal directions are of the same size as the point-like
spread~\cite{Kachelriess:2019taq}. Thus, one will expect
$\sigma\equiv\sigma_{e^+e^-}=\sigma_{pp}/\sqrt{2}$.

The numerical value $\sigma=(1.0\pm 0.1)$\,fm has been shown to reproduce a
wide range of
experimental data on $pp$, $e^+e^-$ and $pN$ collisions, as well as baryonic
femtoscopy, within experimental and theoretical
uncertainties~\cite{Tjemsland:2020bzu}. This value is also in
agreement with the physical interpretation of the model, being thus a strong
indication of the validity of the underlying model assumptions.
The spread should in principle vary between events; in particular, it should
depend on the impact parameter and multiplicity. Moreover,
$\sigma_\perp\lesssim \sigma_\parallel$. With improved experimental data and
improved event generators, one may have therefore to tune $\sigma_\perp$ and
$\sigma_\parallel$ independently and vary them as a function of multiplicity.

\subsection{Spatial correlations in event generators}
\label{sec:pythia}

Some event generators, like Pythia~\cite{Bierlich:2022pfr,Ferreres-Sole:2018vgo}
and EPOS~\cite{Werner:2005jf,Pierog:2013ria}, include a semi-classical
description of the space-time evolution of the cascade. If one employs the
space-time treatment of an event generator, the coalescence weight becomes
\begin{equation}
  w=\mathcal{D}(\vec q,\vec r)=
    3\exp\left\{-\frac{1}{d^2}\left(\vec{r} + \frac{\vec{q}t}{m}\right)^2 -
    q^2d^2\right\},
\end{equation}
and can be extended to a two-Gaussian wave function as Eq.~\eqref{fig:r_core}.
Heisenberg's  uncertainty  relation  limits the precision of the space-time
information a specific event can contain. As a result, the  space-time evolution
predicted by these generators can be only an approximation to the expected
probability distributions. Thus, this approach is  merely a change of the
semi-classical description of the nucleon distribution from that discussed in 
the previous subsection to that supplied by the event generator. It has,
however, some advantages: First, the non-trivial Lorentz transformation of
the emission volume can be taken into account in a straight-forward manner.
For example, one does not have to assume that the momenta of the quark pair
initiating the cascade
are directed along the beam direction. Thus, more complicated processes, like
$\Upsilon\to ggg$, are trivial to consider, provided that the event generator
describes the process accurately. Second,
the emission volume is expected to be strongly correlated with the 
centrality of the collision in $pN$ and $NN$ collisions, and thus the
multiplicity. These effects can in principle be described by an event
generator. Third, a weak energy dependence of the emission volume
is expected. Note that these effects will likely only be visible in
accelerator data, when narrow parts of the phase space are considered. In
cosmic ray physics, however, it is more appropriate to use an event generator
which is specialised to such applications, e.g.,
QGSJET~\cite{Ostapchenko:2010vb,Ostapchenko:2013pia}.

\section{Examples}
\label{sec:examples}

In this section, we will be considering a few examples of antinuclei production
in small interacting systems, using Pythia 8 as event generator.
One should note that the space-time treatment of  parton--parton
interactions in Pythia is not yet complete and there exist  yet no official
tunes~\cite{Ferreres-Sole:2018vgo}. In particular, the geometrical contribution
to the longitudinal spread is not
implemented, i.e., all parton--parton interactions
occur at $z=0$. As such, we cannot expect at this time Pythia to perfectly
reproduce the experimental data. 
Nevertheless, the examples we are considering can be used to tune and
develop Pythia's space-time picture. Moreover, they will highlight some of the important
features of the WiFunC model.

\subsection{Emission volume in $pp$ collisions at LHC}
\label{sec:volume}

The ALICE collaboration measured recently the source radius of the baryon
emission at 13\,TeV in $pp$ collisions by assuming an isotropic Gaussian source
profile in the femtoscopy framework~\cite{Acharya:2020dfb}. As discussed in
section~\ref{sec:femtoscopy}, the  source radius
is directly connected to the Wigner function in the coalescence model via
Eq.~\eqref{eq:main}. Although a simplified description of the source was used,
the treatment of $W_{np}$ in the coalescence model should reproduce this
measurement. It allows us thereby to tune the coalescence model completely
independently of antideuteron measurements.
A caveat is that the measurement is conducted in the lab frame, while the
source is defined in the pair rest frame. Therefore, the measured source size
is the Euclidean distance in the lab frame at ``freeze-out'' boosted into the
pair rest frame. This naturally explains the $m_\mathrm{T}$ scaling
observed in Ref.~\cite{Acharya:2020dfb}.

\begin{figure}
  \hspace*{-0.9cm}
  \includegraphics[width=0.58\textwidth]{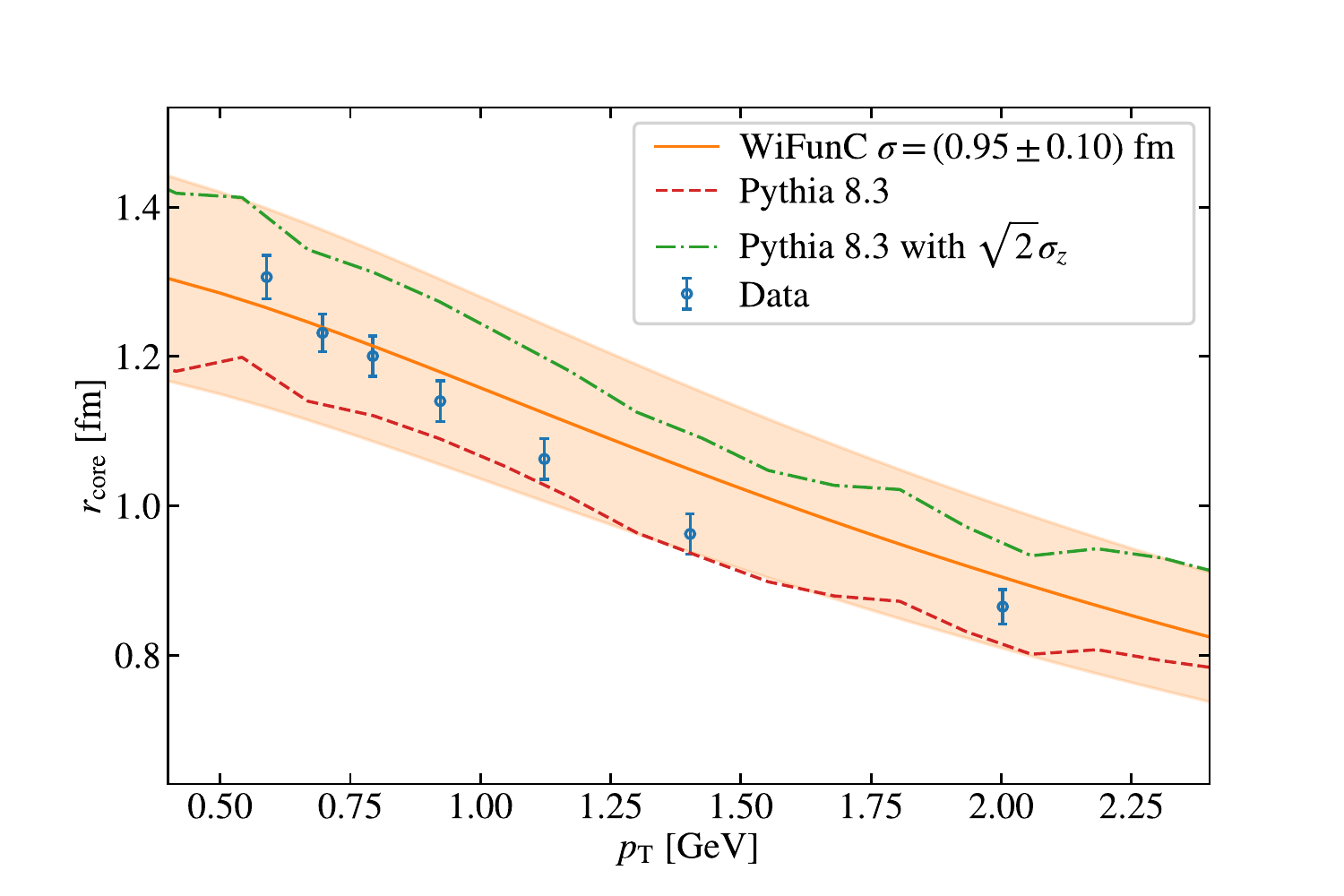}
  \caption{The Gaussian emission size of $pp$ and $\bar{p}\bar{p}$,
  $r_\mathrm{core}$, measured by the ALICE collaboration~\cite{Acharya:2020dfb}
  (blue circles)
  is compared to the prediction of the WiFunC model
   using the space-time picture of Pythia 8 (red dashed line) and the
  Gaussian ansatz for the emission volume~\cite{Kachelriess:2020amp} (orange solid line).
  Since the longitudinal
  geometrical spread is not yet included in Pythia, we show the results
  with a longitudinal spread added by hand (green dashed dotted) for
  visualisation.}
  \label{fig:r_core}
\end{figure}

In Fig.~\ref{fig:r_core}, we compare the $pp\otimes\bar p \bar p$ source radius
measured by ALICE to that predicted by the Gaussian ansatz in the WiFunC
model (see Ref.~\cite{Kachelriess:2020amp}) and by the space-time picture
implemented in Pythia. It is clear that the qualitative behaviour of the $m_T$
scaling is well reproduced by Pythia, while the overall source size is
under-estimated. The latter is expected as the longitudinal geometrical
spread is not yet included in Pythia. For illustration, we have added an
additional line where $\sigma_z$ was increased by a factor $\sqrt{2}$; the
resulting agreement indicates that
the space-time picture in Pythia has the potential to reproduce the
experimental data. In turn, these data can be used to tune the space-time
approach of Pythia.

\subsection{Deuteron spectrum at LHC}
\label{sec:spectrum}

\begin{figure}
  \hspace*{-0.6cm}
  \includegraphics[width=0.53\textwidth]{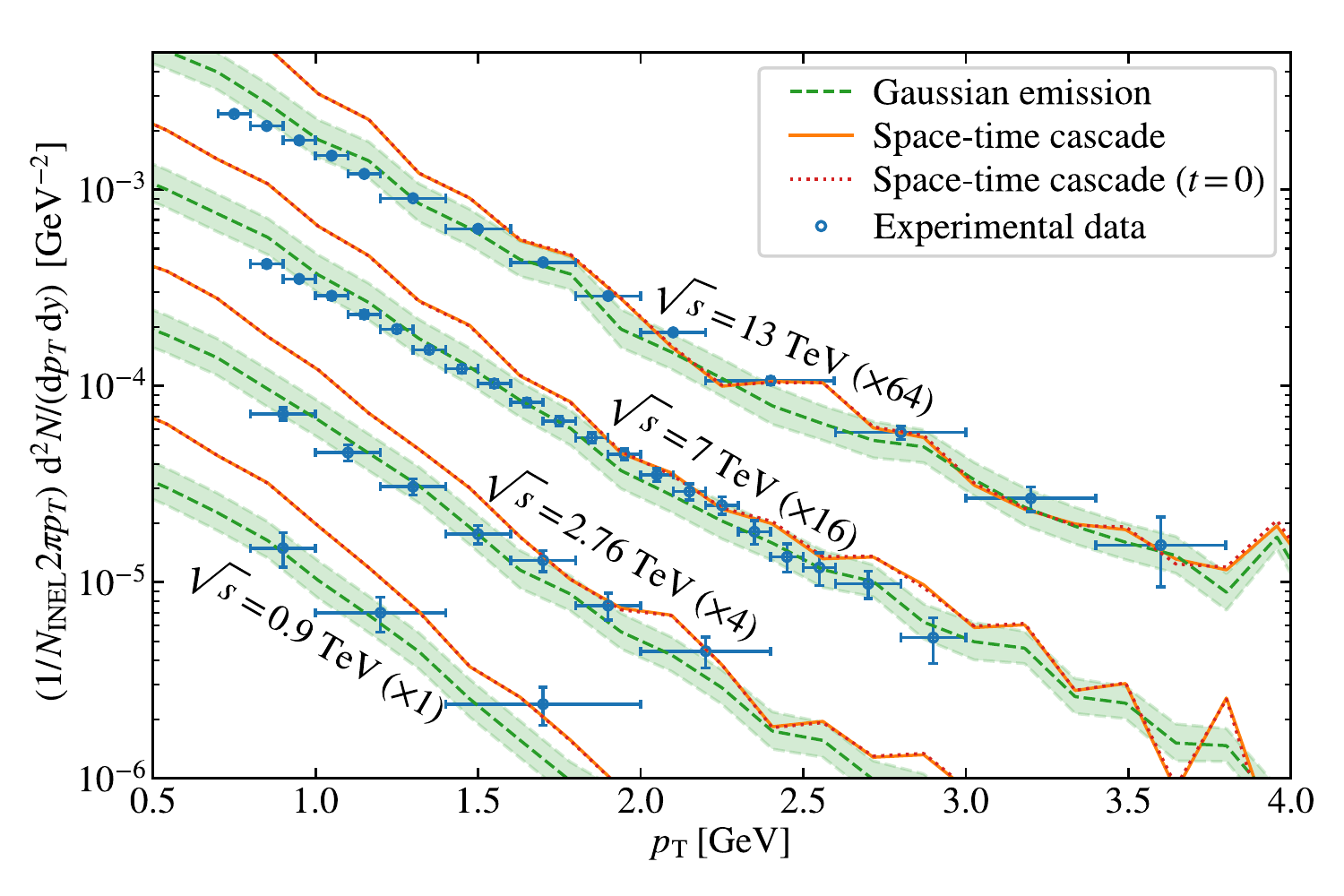}
  \caption{
    The antideuteron spectrum in $pp$ collisions at
    $\sqrt{s}=0.9$, 2.7, 7 and 13\,TeV, predicted by Pythia~8, using a
    Gaussian ansatz with $\sigma=(1.0\pm0.1)\unit{fm}$ (green dashed line)
    and  the space-time approach of
    Pythia (orange line), is compared to the experimental data of
    the ALICE collaboration~\cite{ALICE:2017xrp,Acharya:2020sfy} (the data
    at 13 TeV is multiplied by a factor 0.79 to normalise the spectrum to the
    total number of inelastic events).
    The result without
    the equal-time approximation (red dotted line) is shown for completeness.
  }
    \label{fig:spectrum}
\end{figure}

In Fig.~\ref{fig:spectrum}, the deuteron spectrum in $pp$ collisions at
0.9, 2.76, 7 and 13\,TeV, as predicted by the WiFunC model with the Gaussian
emission volume and with the space-time picture implemented in Pythia,
is compared to the experimental data measured by the ALICE
collaboration~\cite{ALICE:2017xrp,Acharya:2020sfy}.
It is clear that the space-time approach of
Pythia is overproducing antinuclei, as expected from the under-estimated 
longitudinal size 
discussed in the previous subsection.
We note again that these measurements can be used to tune Pythia's space-time
picture. Due to the composite structure of the deuteron, one can also use
antinuclei experiments to tune the event generator to two-particle correlations.

The lines with and without the equal-time approximation completely overlap.
That is, Pythia predicts that the inaccuracy of the equal-time approximation is
neglegible  at LHC energies:
Although the uncertainty in the emission volume in single events is
of order 10\%, the effect is suppressed since the coalescence condition
requires the pairs of nucleons to be close-by in phase space.

\subsection{Enhanced coalescence probability in jets}
\label{sec:jets}

The ALICE collaboration has measured an enhanced (anti-) deuteron coalescence
probability in jets~\cite{ALICE:2022jmr,ALICE:2020hjy}, compared to the
underlying events for $pp$ collisions at 13\;TeV.
More concretely, the measured coalescence factor
\begin{equation}
  B_2 =
  \left(
    \frac{1}{2\pi p_T^\mathrm{deut}}
      \frac{\mathrm{d}^2N_\mathrm{deut}}{\mathrm{d}y\;\mathrm{d}p_T^\mathrm{deut}}
  \right)
  /
  \left(
    \frac{1}{2\pi p_T^p}
      \frac{\mathrm{d}^2N_p}{\mathrm{d}y\;\mathrm{d}p_T^p}
  \right)^2
  \label{eq:B2}
\end{equation}
for $p^p_T = p_T^d/2$ and $|y|<0.5$
is a factor $\approx 10$ larger in a jet than in the underlying event. In the
coalescence model, this is naturally explained by the larger phase space
density of nucleons in the jet, and is therefore a strong indication
that coalescence is a major production mechanism for   deuterons.
Moreover, this experiment may prove useful for understanding the exact nature
of the coalescence mechanism.

\begin{figure}
    \hspace*{-0.9cm}
  \includegraphics[width=0.58\textwidth]{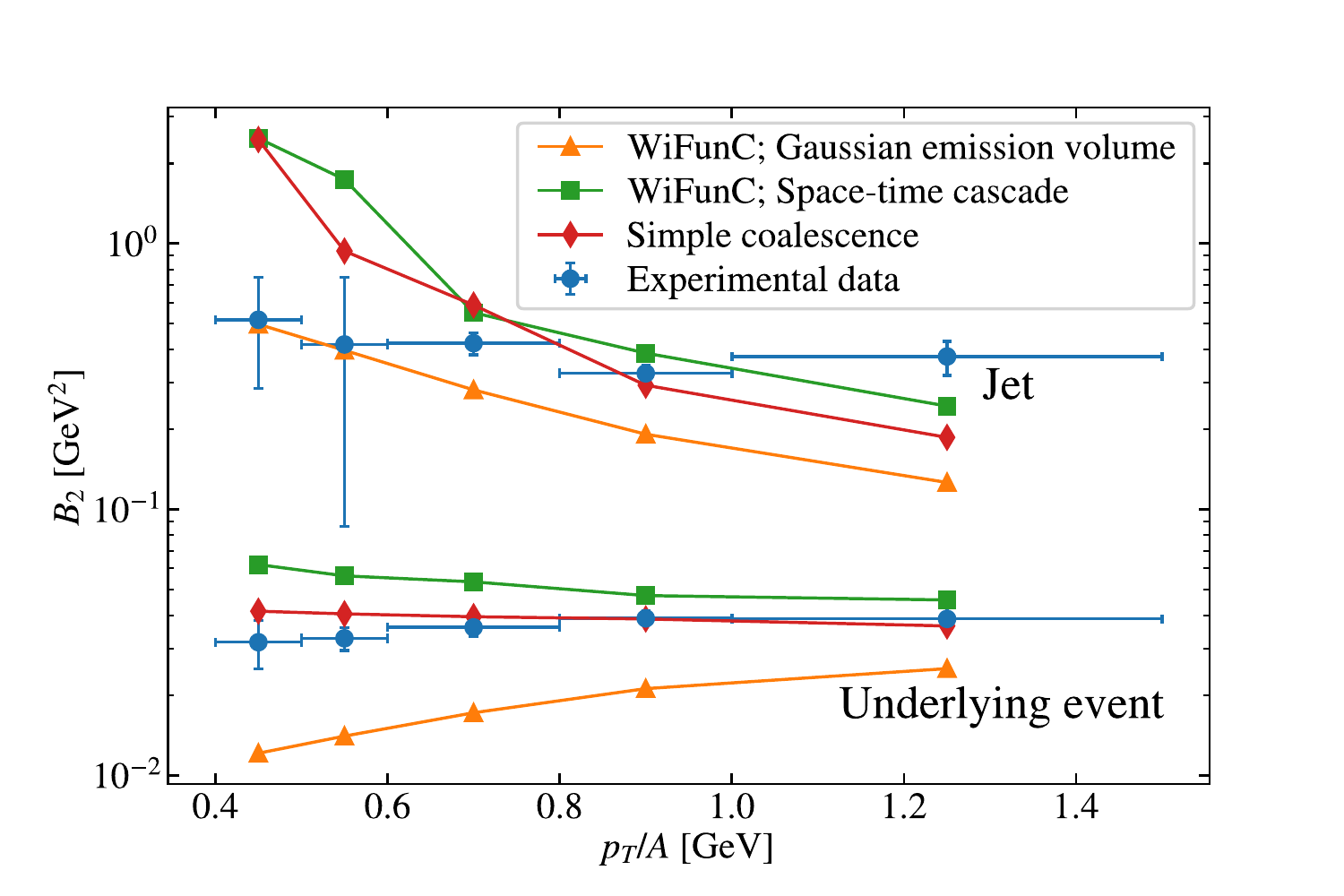}
  \caption{
    The measured coalescence factor $B_2$ in $pp$ collisions at
    $\sqrt{s}=13$\,TeV (blue circles) is compared to the predictions of
    Pythia 8.3 combined with the WiFunC model using a simple Gaussian ansatz
    (orange triangles) or based on the  space-time treatment of  Pythia (green squares). In addition,
    the results using the
    simple coalescence model (red diamonds) are shown
    for comparison.
   }
  \label{fig:jet}
\end{figure}

In Fig.~\ref{fig:jet}, we compare the coalescence factor~\eqref{eq:B2}
predicted by the WiFunC model with a simple Gaussian ansatz (blue) and
using the space-time picture in Pythia (orange).
The results were obtained simulating inelastic $pp$ collisions at 13\,TeV,
using Pythia~8.3 and enforcing the experimental triggers and cuts used in
the event selection~\cite{ALICE:2022jmr,ALICE:2020hjy}.
The jet axis was approximated as the region with an azimuthal
angle $|\Delta \phi| < 60^\circ$ around the so-called leading particle,
as explained in Ref.~\cite{ALICE:2020hjy}. Any
charged particle at midrapidity ($|y|<0.5$) and high transverse
momentum ($p_T> 5\unit{GeV}$) is considered a leading particle.
In the same manner, the underlying event was
approximated by the region $60^\circ < |\Delta \phi | < 120^\circ$.

The overall results shown in Fig.~\ref{fig:jet} are consistent with
those of Ref.~\cite{ALICE:2022jmr}: There is an enhancement of a factor
$\sim 10$
in the coalescence probability (i.e., the coalescence factor $B_2$) in the jet,
compared to the underlying event. For comparison, we also use the simple
coalescence
model~\cite{ALICE:2022jmr} (green)
with a hard cutoff in momentum space, $p<0.285$\,GeV, and a statistical
weight 3/8.
We emphasise  that no fitting was performed, and the result from the WiFunC
model (orange and blue) should be considered as a prediction. In accordance
with Fig.~\ref{fig:r_core}, the space-time treatment overpredicts the
coalescence probability.
One should further note that the emission volume used in the
simple Gaussian ansatz includes a Lorentz transformation relative to the beam
axis, which is expected to be a valid approximation for a typical $pp$
interaction. However, in Fig.~\ref{fig:jet}, we are only considering events
within a clear jet, in which case the boost should be done relative to the
initial parton in the parton cascades. This is one of the main perks in
using the space-time treatment in Pythia, since more complicated geometries
are automatically taken into account.

\subsection{Energy dependence of the emission volume}
\label{sec:energy}

The emission volume is expected to have a weak energy
dependence~\cite{Kachelriess:2019taq}.  Within the current experimental and
theoretical uncertainties, the emission volume is consistent with being
constant~\cite{Tjemsland:2020bzu}. The expected energy dependence and its
relevance to coalescence is however not trivial: At high energies, the source
size measured via  femtoscopy experiments will increase and be much
larger than 1\,fm. For instance, the average $\Gamma$ factor of nucleons
in their pair rest-frame increases with the center-of-mass energy $\sqrt{s}$
of the collision. As a result, the hadronisation length
$\ell\simeq \Gamma/ m_N$ increases with $\sqrt{s}$. This growth will
affect mainly the longitudinal emission length.
Moreover, multiple scattering  in hadronic collisions enlarges the
source volume additionally. While the first effect is strongly suppressed
in the production of light nuclei because of the coalesence weight $w$,
it is also suppressed in femtospectroscopy measurements because of experimental
cuts. For instance, the ALICE collaboration used $q<0.375$\,GeV, in
addition to the trigger condition and the rapidity cut.

\begin{figure}
    \hspace*{-0.9cm}
  \includegraphics[width=0.58\textwidth]{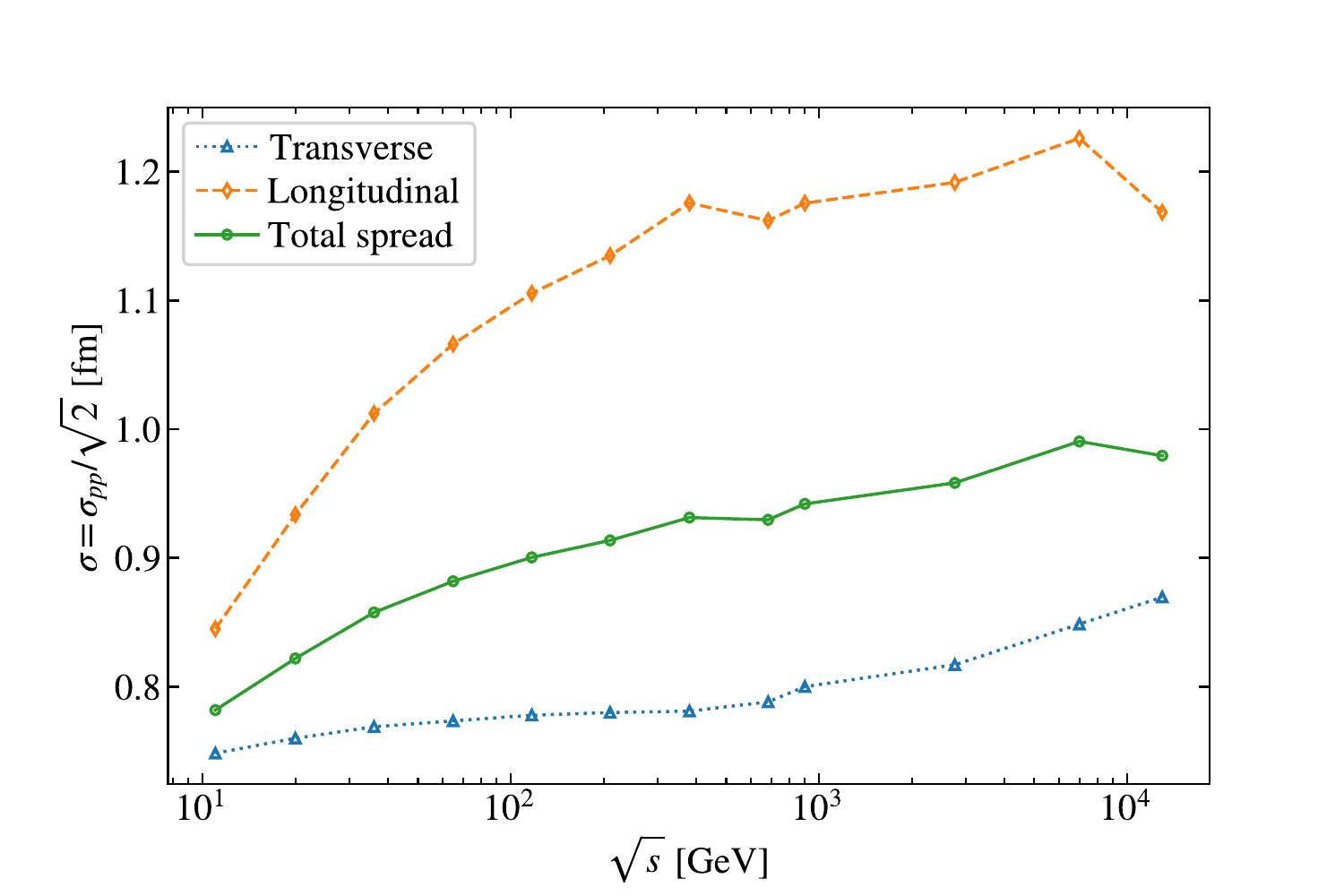}
  \caption{
    The spread $\sigma$ predicted by Pythia is computed as the 
    rms value of the size of the nucleon emission region,
     while using the weight $\exp(-d^2q^2)$.
    The transverse (blue triangles), longitudinal (orange diamonds) and 
    total (green circles) spreads are plotted as a function of the center of mass energy in 
    $pp$ collisions. }
  \label{fig:energy}
\end{figure}

In order to test this expectation and at the same time to highlight some
differences between the Gaussian ansatz for the emission volume and the
space-time picture of the event generator, we plot in Fig.~\ref{fig:energy}
the predicted energy  dependence of $\sigma$ by Pythia~8.3 in $pp$ collisions.
A weight $\exp(-d^2q^2)$ was included
($q$ being the nucleon momentum in the pair rest frame)
to highlight the ``coalescence relevant'' source size.
Pythia predicts, as expected~\cite{Kachelriess:2019taq}, a weak energy
dependence of the emission volume and $\sigma_\parallel>\sigma_\perp$.
The energy dependence can be explained by correlations between the position
and momentum:
Initially, the spread $\sigma$ increases due to the increased energy available
to the nucleons; the increase is dominated by nucleons produced back-to-back.
At some point, the cut in momentum space suppresses the
emission volume, making $\sigma$ approximately constant.

Furthermore, there is a significant difference in the energy evolution of the
spread in the longitudinal and transverse direction.
In the Gaussian ansatz of the source
volume [cf. with Eq.~\eqref{eq:zeta}],
the longitudinal spread will be constant while the transverse spread
will effectively be Lorentz contracted for large transverse momenta:
$\sigma_\perp = \sigma m/m_T$. Meanwhile, using Pythia, the Lorentz
boost is performed on a pair-by-pair basis and is thus not defined relative to
the initial particle beam. Therefore, the expected transverse contraction
in Pythia will occur  both for $\sigma_\perp$ and $\sigma_\parallel$.

Cosmic ray antinuclei are mainly produced by primary protons colliding with the
interstellar medium at energies 10--20$\unit{GeV}$ in the center-of-mass
frame. According to the results in
Fig.~\ref{fig:energy}, Pythia predicts a decrease
of $\sigma$ by $\sim 0.1 \unit{fm}$, when moving from LHC to such low
energies.
Closer to the threshold, outside the validity range of Pythia, anti-correlations
can increase the baryon emission volume but will have little impact on the
final deuteron spectrum since the nuclei are already suppressed by the
anti-correlations in momentum. This will in any case have negligible effects on
cosmic ray studies.

\subsection{$\Upsilon$ decays}
\label{sec:upsilon}

The decay of $\Upsilon$ is interesting because one can learn about the
hadronisation and coalescence process at low energies. 
Recently, Ref.~\cite{Marietti:2022jil} systematically tested phase-space Monte
Carlo models
on $\Upsilon$ decay data. Using the WiFunC model, it was found that the emission
size $\sigma\simeq 1.6$\,fm---greatly larger than the expected
$\simeq 1$\,fm---is needed to reproduce the measured antideuteron yield.
This may have three explanations~\cite{Kachelriess:2022khq}:
(1) the WiFunC model fails, 
(2) the event generator over-predicts the nucleon yield or nucleon correlations, or
(3) the nucleon emission volume is larger than expected in this process.

To test the first explanations,
we simulate the decay of $10^7$ $\Upsilon$ using Pythia~8.3, turning off the
decay of strong resonances.
In the WiFunC model with the Gaussian ansatz, we obtain\footnote{
  Due to the lack of a preferred direction,
  we neglect the Lorentz boost in the transverse direction.
}
$B(\Upsilon\to \bar dX)=6.7^{+0.1}_{-0.2}\times 10^{-5}$ with
$\sigma=(1.0\pm0.1)\unit{fm}$. In agreement with Ref.~\cite{Marietti:2022jil},
we need $\sigma\simeq 1.5$\,fm to reproduce the value measured by
BaBar~\cite{BaBar:2014ssg}, $2.81\pm 0.49^{+0.2}_{-0.24}$.
Using  the space-time treatment of Pythia, we obtain
$B(\Upsilon\to \bar dX)=18.0\times 10^{-5}$, and an effective size
$\sigma=0.83$\,fm. Without the equal time approximation, the result
is $B(\Upsilon \to \bar dX)= 17.2\times 10^{-5}$ with an effective size
$\sigma=0.93$\,fm. This is a change of 4.5\%. Even if the estimated
emission volume in Pythia is similar to the one used in the Gaussian ansatz,
the branching ratio is a factor 2--3 larger, indicating a substantial
enhancement due to position and momentum correlations. In all cases, the
WiFunC model over-predicts the measurement, which may well be due
to uncertainties in the event generator.

In order to test the hypothesis that Pythia over-predicts the nucleon
yield in the meson-to-three-gluon decay\footnote{
  In the decay tables in Pythia,
  the $J/\psi$ meson decays mainly into
  two gluons, even though the dominant decay channel is 
  $J/\psi\to ggg$~\cite{ParticleDataGroup:2022pth}. We therefore change
  this decay channel to three gluons, like for
  $\Upsilon$, in the simuations.
}, we simulate the decay of $J/\psi$ and compare the measured branching
ratio~\cite{ParticleDataGroup:2022pth} of common decays into nucleons and
pions\footnote{
  We neglect the contributions from resonances, as well as
  final state photons since Pythia includes Bremstralung photons in the decays.
}.
The result is shown in Tab.~\ref{tab:decay}. As readily seen from the table, 
Pythia has a tendency to under-estimate the branching ratio into pions, and
to over-estimate the branching into nucleons. This is a strong indication that
Pythia over-predicts the nucleon production in $J/\psi$, and thus
$\Upsilon$ decays. The nucleon yield is overproduced by a factor 2--3, implying
that the deuteron yield may be overestimated by a factor 4--9.

In Pythia, the $\Upsilon$ meson decays
mainly into three gluons, which may initiate parton showers and hadronise. 
In a different line of thought, the three gluons expand a triangular
Lund string, and so the hadronic emission
length might be substantially larger than in other processes,
$\sim 3\unit{fm}$~\cite{Gustafson:1993mm}.

In conclusion, the theoretical uncertainties prevent at present a conclusion
about the size of the emission volume in $\Upsilon$ decays.
The baryonic production and baryon--baryon femtoscopy measurements in
$\Upsilon$ decays are therefore highly warranted.
This will allow one, in tandem with the
antideuteron data, to learn about hadronic meson decays, the
hadronisation process and the coalescence process. Moreover, it may increase
significantly the predictive power for some exotic antinuclei production
mechanisms, such as dark matter decays or annihilation.

\begin{table}
  \centering
\begin{tabular}{c c c}
  \hline
  Decay & Measured value~\cite{ParticleDataGroup:2022pth} & Pythia \\
  \hline
  \hline
  $2(\pi^+\pi^-)\pi^0$ & $(3.71\pm0.28)\times 10^{-2}$ & $2.50\times 10^{-3}$\\
  $3(\pi^+\pi^-)\pi^0$ & $(2.9\pm0.6)\times 10^{-2}$   & $0$\\
  $\pi^+\pi^-3\pi^0$ & $(1.9\pm0.9)\times 10^{-2}$     & $1.36\times 10^{-3}$\\
  $\pi^+\pi^-4\pi^0$ & $(6.1\pm1.3)\times 10^{-3}$     & $6.5\times 10^{-5}$\\
  $\pi^+\pi^-\pi^0$ & $(2.10\pm0.08)\times 10^{-2}$    & $1.51\times 10^{-2}$\\
  $2(\pi^+\pi^-\pi^0)$ & $(1.61\pm0.20)\times 10^{-2}$ & $2.50\times 10^{-4}$\\
  $\pi^+\pi^-\pi^0K^+K^-$ & $(1.20\pm0.30)\times 10^{-2}$ & $6.06\times 10^{-3}$\\
  $\pi^+\pi^-$ & $(1.47\pm0.14)\times 10^{-4}$ & $7.68\times 10^{-3}$\\ 
  $2(\pi^+\pi^-)$ & $(3.57\pm0.30)\times 10^{-3}$ & $5.31\times 10^{-3}$\\
  $\gamma2\pi^+2\pi^-$ & $(2.8\pm0.5)\times 10^{-3}$ & -- \\
  $3(\pi^+\pi^-)$ & $(4.3\pm0.4)\times 10^{-3}$ & $5.90\times 10^{-5}$\\
  $2(\pi^+\pi^-)3\pi^0$ & $(6.2\pm0.9)\times 10^{-2}$ & $7.30\times 10^{-6}$\\
  $4(\pi^+\pi^-)\pi^0$ & $(9.0\pm 3.0)\times 10^{-3}$ & $0$\\
  \hline
  Total & $0.222 \pm 0.015$ & 0.057 \\
  \hline
  $p \bar p$ & $(2.120\pm0.029)\times 10^{-3}$    & $1.36\times 10^{-2}$\\
  $p \bar p\pi^0$ & $(1.19\pm0.08)\times 10^{-3}$    & $4.31\times 10^{-3}$\\
  $p \bar p\pi^+\pi^-$ & $(6.0\pm0.5)\times 10^{-3}$    & $9.07\times 10^{-4}$\\
  $p \bar p\pi^+\pi^-\pi^0$ & $(2.3\pm0.9)\times 10^{-3}$&$1.02\times 10^{-4}$\\
  $p \bar n\pi^-$ & $(2.12\pm0.09)\times 10^{-3}$    & $7.47\times 10^{-3}$\\
  $n \bar n$ & $(2.09\pm0.16)\times 10^{-3}$    & $1.36\times 10^{-2}$\\
  \hline
  Total & $(1.58\pm0.01)\times 10^{-2}$ & $4.10\times 10^{-2}$ \\
  \hline
\end{tabular}
  \caption{Branching ratios of common $J/\psi$ decays into pions and nucleons.}
  \label{tab:decay}
\end{table}

\section{Summary and conclusions}

We have discussed the WiFunC model, a coalescence model that allows
one to include  momentum and spatial correlations on an event-by-event basis.
Two choices for the nucleon emission volume were discussed: (1) a
Gaussian ansatz, and (2) using the emission volume provided by an event
generator. In the latter case, one can go beyond the equal-time approximation
which until now has been invariably assumed. We have shown that this
approximation leads to a $\mathcal{O}(10\unit{\%})$
uncertainty in the coalescence probability in processes close to the production
threshold, such as $\Upsilon$ decays. The error is strongly reduced at high energies, implying that non-equal production times can be neglected
for hadronic collisions at LHC.

As concrete examples, we considered $pp$ collisions and $\Upsilon$ decays, 
using Pythia 8. The Gaussian ansatz for the emission volume leads to a
satisfactory description of the baryon emission volume and the antideuteron
spectrum measured by the ALICE collaboration at LHC, while overpredicting the
antideuteron production in  $\Upsilon$ decays. We have argued, based on
experimental data on nucleon production in $J/\psi$ decays,
that Pythia likely overpredicts the nucleon production in $\Upsilon$ decays.
The space-time approach of Pythia~8, on the other hand,
underpredicts the nucleon emission volume and fails to accurately
describe the antideuteron spectrum. However,  these deficiencies are most
likely  explained by the fact that the space-time treatment is not yet complete
and has not yet been tuned to experimental data. Importantly, this implies
that the coalescence framework introduced in this work can be used to tune
the space-time  treatments and momentum correlations in event generators, when comparing
them to antideuteron  and femtoscopy data. Once nucleon production in $\Upsilon$
decays is measured, one can use also antinuclei to probe the hadronisation
process. In addition, we predicted the energy dependence of the emission
volume using Pythia 8. This resulted, as expected, in a weak energy dependence,
consistent with a constant $(1.0\pm 0.1)$\,fm within experimental and
theoretical uncertainties.

This work has been motivated by an increasing amount of high-precision
data on antinuclei production in small interacting systems, obtained by, e.g.,
the ALICE, NA61/SHINE and BELLE-II experiments. Our framework paves the way
for using these antinuclei measurements to tune the space-time picture and
momentum correlations in   event generators used to describe these data.
Improving thereby the {accuracy of such generators, regarding the 
description of antinuclei production,
may furthermore have an important impact on predictions of antinuclei
production by cosmic rays and dark matter.

\vspace*{0.6cm}\noindent
{\bf Note added:}
While finalising this manuscript, the related work~\cite{Horst:2023oti}
appeared on the arxiv. The authors of that work employ the
equal-time approximation together with Eq.~(\ref{eq:p}) and a Gaussian
ansatz for $H_{np}$. The width of the Gaussian was however treated as
a variable, $\sigma\to\sigma(\vec{r}, t)$, what is inconsistent with
the assumptions needed to derive the deuteron yield in Eq.~\eqref{eq:p}.
We also note that our results for the emission volume, based on
 Pythia~8, are in disagreement with theirs: We obtain with
Pythia a source size which decreases with transverse mass---in agreement with
the experimental data---while the source size derived in
Ref.~\cite{Horst:2023oti} increases.
This discrepency is likely mainly caused by a different  interpretation of
the effect of the equal-time approximation in a femtoscopy experiment:
In deriving the core size shown in Fig.~\ref{fig:r_core}, we enforce $t=0$
in the lab frame. Meanwhile, Ref.~\cite{Horst:2023oti} enlarges
the emission volume by propagating the produced particles until $t=0$.

\section{Acknowledgements}

This work benefited from discussions at the workshop ``Antinuclei in the
Universe''  at the Munich Institute for Astro- and Particle
Physics (MIAPP) which is funded by the Deutsche Forschungsgemeinschaft (DFG,
German Research Foundation) under Germany's Excellence Strategy -- EXC-2094 --
390783311. S.O. acknowledges support from the Deutsche
Forschungsgemeinschaft (project number 465275045).


\begin{thebibliography}{65}
\expandafter\ifx\csname natexlab\endcsname\relax\def\natexlab#1{#1}\fi
\expandafter\ifx\csname bibnamefont\endcsname\relax
  \def\bibnamefont#1{#1}\fi
\expandafter\ifx\csname bibfnamefont\endcsname\relax
  \def\bibfnamefont#1{#1}\fi
\expandafter\ifx\csname citenamefont\endcsname\relax
  \def\citenamefont#1{#1}\fi
\expandafter\ifx\csname url\endcsname\relax
  \def\url#1{\texttt{#1}}\fi
\expandafter\ifx\csname urlprefix\endcsname\relax\def\urlprefix{URL }\fi
\providecommand{\bibinfo}[2]{#2}
\providecommand{\eprint}[2][]{\url{#2}}

\bibitem[{\citenamefont{Caines}(2017)}]{Caines:2017vvs}
\bibinfo{author}{\bibfnamefont{H.}~\bibnamefont{Caines}},
  \bibinfo{journal}{Nucl. Phys.} \textbf{\bibinfo{volume}{A967}},
  \bibinfo{pages}{121} (\bibinfo{year}{2017}).

\bibitem[{\citenamefont{Chardonnet et~al.}(1997)\citenamefont{Chardonnet,
  Orloff, and Salati}}]{Chardonnet:1997dv}
\bibinfo{author}{\bibfnamefont{P.}~\bibnamefont{Chardonnet}},
  \bibinfo{author}{\bibfnamefont{J.}~\bibnamefont{Orloff}}, \bibnamefont{and}
  \bibinfo{author}{\bibfnamefont{P.}~\bibnamefont{Salati}},
  \bibinfo{journal}{Phys. Lett.} \textbf{\bibinfo{volume}{B409}},
  \bibinfo{pages}{313} (\bibinfo{year}{1997}), \eprint{astro-ph/9705110}.

\bibitem[{\citenamefont{Donato et~al.}(2000)\citenamefont{Donato, Fornengo, and
  Salati}}]{Donato:1999gy}
\bibinfo{author}{\bibfnamefont{F.}~\bibnamefont{Donato}},
  \bibinfo{author}{\bibfnamefont{N.}~\bibnamefont{Fornengo}}, \bibnamefont{and}
  \bibinfo{author}{\bibfnamefont{P.}~\bibnamefont{Salati}},
  \bibinfo{journal}{Phys. Rev.} \textbf{\bibinfo{volume}{D62}},
  \bibinfo{pages}{043003} (\bibinfo{year}{2000}), \eprint{hep-ph/9904481}.

\bibitem[{\citenamefont{von Doetinchem et~al.}(2020)}]{vonDoetinchem:2020vbj}
\bibinfo{author}{\bibfnamefont{P.}~\bibnamefont{von Doetinchem}}
  \bibnamefont{et~al.}, \bibinfo{journal}{JCAP} \textbf{\bibinfo{volume}{08}},
  \bibinfo{pages}{035} (\bibinfo{year}{2020}), \eprint{2002.04163}.

\bibitem[{\citenamefont{Csernai and Kapusta}(1986)}]{Csernai:1986qf}
\bibinfo{author}{\bibfnamefont{L.}~\bibnamefont{Csernai}} \bibnamefont{and}
  \bibinfo{author}{\bibfnamefont{J.~I.} \bibnamefont{Kapusta}},
  \bibinfo{journal}{Phys. Rept.} \textbf{\bibinfo{volume}{131}},
  \bibinfo{pages}{223} (\bibinfo{year}{1986}).

\bibitem[{\citenamefont{Nagle et~al.}(1996)\citenamefont{Nagle, Kumar,
  Kusnezov, Sorge, and Mattiello}}]{Nagle:1996vp}
\bibinfo{author}{\bibfnamefont{J.~L.} \bibnamefont{Nagle}},
  \bibinfo{author}{\bibfnamefont{B.~S.} \bibnamefont{Kumar}},
  \bibinfo{author}{\bibfnamefont{D.}~\bibnamefont{Kusnezov}},
  \bibinfo{author}{\bibfnamefont{H.}~\bibnamefont{Sorge}}, \bibnamefont{and}
  \bibinfo{author}{\bibfnamefont{R.}~\bibnamefont{Mattiello}},
  \bibinfo{journal}{Phys. Rev. C} \textbf{\bibinfo{volume}{53}},
  \bibinfo{pages}{367} (\bibinfo{year}{1996}).

\bibitem[{\citenamefont{Schwarzschild and
  Zupancic}(1963)}]{Schwarzschild:1963zz}
\bibinfo{author}{\bibfnamefont{A.}~\bibnamefont{Schwarzschild}}
  \bibnamefont{and} \bibinfo{author}{\bibfnamefont{C.}~\bibnamefont{Zupancic}},
  \bibinfo{journal}{Phys. Rev.} \textbf{\bibinfo{volume}{129}},
  \bibinfo{pages}{854} (\bibinfo{year}{1963}).

\bibitem[{\citenamefont{Butler and Pearson}(1963)}]{butler_deuterons_1963}
\bibinfo{author}{\bibfnamefont{S.~T.} \bibnamefont{Butler}} \bibnamefont{and}
  \bibinfo{author}{\bibfnamefont{C.~A.} \bibnamefont{Pearson}},
  \bibinfo{journal}{Phys. Rev.} \textbf{\bibinfo{volume}{129}},
  \bibinfo{pages}{836} (\bibinfo{year}{1963}), ISSN \bibinfo{issn}{0031-899X}.

\bibitem[{\citenamefont{Gustafson and Hakkinen}(1994)}]{Gustafson:1993mm}
\bibinfo{author}{\bibfnamefont{G.}~\bibnamefont{Gustafson}} \bibnamefont{and}
  \bibinfo{author}{\bibfnamefont{J.}~\bibnamefont{Hakkinen}},
  \bibinfo{journal}{Z. Phys. C} \textbf{\bibinfo{volume}{61}},
  \bibinfo{pages}{683} (\bibinfo{year}{1994}).

\bibitem[{\citenamefont{Dal}(2011)}]{Dal_thesis}
\bibinfo{author}{\bibfnamefont{L.~A.} \bibnamefont{Dal}},
  \bibinfo{type}{Master's thesis}, \bibinfo{school}{NTNU Trondheim, available
  at \url{http://hdl.handle.net/11250/246403}} (\bibinfo{year}{2011}),
  \urlprefix\url{http://hdl.handle.net/11250/246403}.

\bibitem[{\citenamefont{Kadastik et~al.}(2010)\citenamefont{Kadastik, Raidal,
  and Strumia}}]{kadastik_enhanced_2010}
\bibinfo{author}{\bibfnamefont{M.}~\bibnamefont{Kadastik}},
  \bibinfo{author}{\bibfnamefont{M.}~\bibnamefont{Raidal}}, \bibnamefont{and}
  \bibinfo{author}{\bibfnamefont{A.}~\bibnamefont{Strumia}},
  \bibinfo{journal}{Phys. Lett.} \textbf{\bibinfo{volume}{B683}},
  \bibinfo{pages}{248} (\bibinfo{year}{2010}), \eprint{0908.1578}.

\bibitem[{\citenamefont{Kachelrie{\ss}
  et~al.}(2020{\natexlab{a}})\citenamefont{Kachelrie{\ss}, Ostapchenko, and
  Tjemsland}}]{Kachelriess:2019taq}
\bibinfo{author}{\bibfnamefont{M.}~\bibnamefont{Kachelrie{\ss}}},
  \bibinfo{author}{\bibfnamefont{S.}~\bibnamefont{Ostapchenko}},
  \bibnamefont{and}
  \bibinfo{author}{\bibfnamefont{J.}~\bibnamefont{Tjemsland}},
  \bibinfo{journal}{Eur. Phys. J.} \textbf{\bibinfo{volume}{A56}},
  \bibinfo{pages}{4} (\bibinfo{year}{2020}{\natexlab{a}}), \eprint{1905.01192}.

\bibitem[{\citenamefont{Kachelrie{\ss}
  et~al.}(2020{\natexlab{b}})\citenamefont{Kachelrie{\ss}, Ostapchenko, and
  Tjemsland}}]{Kachelriess:2020uoh}
\bibinfo{author}{\bibfnamefont{M.}~\bibnamefont{Kachelrie{\ss}}},
  \bibinfo{author}{\bibfnamefont{S.}~\bibnamefont{Ostapchenko}},
  \bibnamefont{and}
  \bibinfo{author}{\bibfnamefont{J.}~\bibnamefont{Tjemsland}},
  \bibinfo{journal}{JCAP} \textbf{\bibinfo{volume}{08}}, \bibinfo{pages}{048}
  (\bibinfo{year}{2020}{\natexlab{b}}), \eprint{2002.10481}.

\bibitem[{\citenamefont{Kachelrie{\ss}
  et~al.}(2021)\citenamefont{Kachelrie{\ss}, Ostapchenko, and
  Tjemsland}}]{Kachelriess:2020amp}
\bibinfo{author}{\bibfnamefont{M.}~\bibnamefont{Kachelrie{\ss}}},
  \bibinfo{author}{\bibfnamefont{S.}~\bibnamefont{Ostapchenko}},
  \bibnamefont{and}
  \bibinfo{author}{\bibfnamefont{J.}~\bibnamefont{Tjemsland}},
  \bibinfo{journal}{Eur. Phys. J. A} \textbf{\bibinfo{volume}{57}},
  \bibinfo{pages}{167} (\bibinfo{year}{2021}), \eprint{2012.04352}.

\bibitem[{\citenamefont{Tjemsland}(2021)}]{Tjemsland:2020bzu}
\bibinfo{author}{\bibfnamefont{J.}~\bibnamefont{Tjemsland}},
  \bibinfo{journal}{PoS} \textbf{\bibinfo{volume}{TOOLS2020}},
  \bibinfo{pages}{006} (\bibinfo{year}{2021}), \eprint{2012.12252}.

\bibitem[{\citenamefont{Kachelrie{\ss}
  et~al.}(2023)\citenamefont{Kachelrie{\ss}, Ostapchenko, and
  Tjemsland}}]{Kachelriess:2022khq}
\bibinfo{author}{\bibfnamefont{M.}~\bibnamefont{Kachelrie{\ss}}},
  \bibinfo{author}{\bibfnamefont{S.}~\bibnamefont{Ostapchenko}},
  \bibnamefont{and}
  \bibinfo{author}{\bibfnamefont{J.}~\bibnamefont{Tjemsland}},
  \bibinfo{journal}{Comput. Phys. Commun.} \textbf{\bibinfo{volume}{287}},
  \bibinfo{pages}{108698} (\bibinfo{year}{2023}), \eprint{2206.00998}.

\bibitem[{\citenamefont{\v{S}erk\v{s}nyt\.{e}
  et~al.}(2022)}]{Serksnyte:2022onw}
\bibinfo{author}{\bibfnamefont{L.}~\bibnamefont{\v{S}erk\v{s}nyt\.{e}}}
  \bibnamefont{et~al.}, \bibinfo{journal}{Phys. Rev. D}
  \textbf{\bibinfo{volume}{105}}, \bibinfo{pages}{083021}
  (\bibinfo{year}{2022}), \eprint{2201.00925}.

\bibitem[{\citenamefont{Scheibl and Heinz}(1999)}]{scheibl_coalescence_1999}
\bibinfo{author}{\bibfnamefont{R.}~\bibnamefont{Scheibl}} \bibnamefont{and}
  \bibinfo{author}{\bibfnamefont{U.~W.} \bibnamefont{Heinz}},
  \bibinfo{journal}{Phys. Rev.} \textbf{\bibinfo{volume}{C59}},
  \bibinfo{pages}{1585} (\bibinfo{year}{1999}), \eprint{nucl-th/9809092}.

\bibitem[{\citenamefont{Sun et~al.}(2018)\citenamefont{Sun, Chen, Ko, Pu, and
  Xu}}]{Sun:2018jhg}
\bibinfo{author}{\bibfnamefont{K.-J.} \bibnamefont{Sun}},
  \bibinfo{author}{\bibfnamefont{L.-W.} \bibnamefont{Chen}},
  \bibinfo{author}{\bibfnamefont{C.~M.} \bibnamefont{Ko}},
  \bibinfo{author}{\bibfnamefont{J.}~\bibnamefont{Pu}}, \bibnamefont{and}
  \bibinfo{author}{\bibfnamefont{Z.}~\bibnamefont{Xu}}, \bibinfo{journal}{Phys.
  Lett. B} \textbf{\bibinfo{volume}{781}}, \bibinfo{pages}{499}
  (\bibinfo{year}{2018}), \eprint{1801.09382}.

\bibitem[{\citenamefont{Sun et~al.}(2017)\citenamefont{Sun, Chen, Ko, and
  Xu}}]{Sun:2017xrx}
\bibinfo{author}{\bibfnamefont{K.-J.} \bibnamefont{Sun}},
  \bibinfo{author}{\bibfnamefont{L.-W.} \bibnamefont{Chen}},
  \bibinfo{author}{\bibfnamefont{C.~M.} \bibnamefont{Ko}}, \bibnamefont{and}
  \bibinfo{author}{\bibfnamefont{Z.}~\bibnamefont{Xu}}, \bibinfo{journal}{Phys.
  Lett. B} \textbf{\bibinfo{volume}{774}}, \bibinfo{pages}{103}
  (\bibinfo{year}{2017}), \eprint{1702.07620}.

\bibitem[{\citenamefont{Shao et~al.}(2020)\citenamefont{Shao, Chen, Ko, and
  Sun}}]{Shao:2019xpj}
\bibinfo{author}{\bibfnamefont{T.}~\bibnamefont{Shao}},
  \bibinfo{author}{\bibfnamefont{J.}~\bibnamefont{Chen}},
  \bibinfo{author}{\bibfnamefont{C.~M.} \bibnamefont{Ko}}, \bibnamefont{and}
  \bibinfo{author}{\bibfnamefont{K.-J.} \bibnamefont{Sun}},
  \bibinfo{journal}{Phys. Lett. B} \textbf{\bibinfo{volume}{801}},
  \bibinfo{pages}{135177} (\bibinfo{year}{2020}), \eprint{1910.14281}.

\bibitem[{\citenamefont{Sun and Ko}(2021)}]{Sun:2020uoj}
\bibinfo{author}{\bibfnamefont{K.-J.} \bibnamefont{Sun}} \bibnamefont{and}
  \bibinfo{author}{\bibfnamefont{C.~M.} \bibnamefont{Ko}},
  \bibinfo{journal}{Phys. Rev. C} \textbf{\bibinfo{volume}{103}},
  \bibinfo{pages}{064909} (\bibinfo{year}{2021}), \eprint{2005.00182}.

\bibitem[{\citenamefont{Zhao et~al.}(2021)\citenamefont{Zhao, Sun, Ko, and
  Luo}}]{Zhao:2021dka}
\bibinfo{author}{\bibfnamefont{W.}~\bibnamefont{Zhao}},
  \bibinfo{author}{\bibfnamefont{K.-j.} \bibnamefont{Sun}},
  \bibinfo{author}{\bibfnamefont{C.~M.} \bibnamefont{Ko}}, \bibnamefont{and}
  \bibinfo{author}{\bibfnamefont{X.}~\bibnamefont{Luo}},
  \bibinfo{journal}{Phys. Lett. B} \textbf{\bibinfo{volume}{820}},
  \bibinfo{pages}{136571} (\bibinfo{year}{2021}), \eprint{2105.14204}.

\bibitem[{\citenamefont{Shao et~al.}(2022)\citenamefont{Shao, Chen, Ma, and
  Xu}}]{Shao:2022eyd}
\bibinfo{author}{\bibfnamefont{T.}~\bibnamefont{Shao}},
  \bibinfo{author}{\bibfnamefont{J.}~\bibnamefont{Chen}},
  \bibinfo{author}{\bibfnamefont{Y.-G.} \bibnamefont{Ma}}, \bibnamefont{and}
  \bibinfo{author}{\bibfnamefont{Z.}~\bibnamefont{Xu}}, \bibinfo{journal}{Phys.
  Rev. C} \textbf{\bibinfo{volume}{105}}, \bibinfo{pages}{065801}
  (\bibinfo{year}{2022}), \eprint{2205.13626}.

\bibitem[{\citenamefont{Gyulassy et~al.}(1983)\citenamefont{Gyulassy, Frankel,
  and Remler}}]{Gyulassy:1982pe}
\bibinfo{author}{\bibfnamefont{M.}~\bibnamefont{Gyulassy}},
  \bibinfo{author}{\bibfnamefont{K.}~\bibnamefont{Frankel}}, \bibnamefont{and}
  \bibinfo{author}{\bibfnamefont{E.~a.} \bibnamefont{Remler}},
  \bibinfo{journal}{Nucl. Phys. A} \textbf{\bibinfo{volume}{402}},
  \bibinfo{pages}{596} (\bibinfo{year}{1983}).

\bibitem[{\citenamefont{Danielewicz and Bertsch}(1991)}]{Danielewicz:1991dh}
\bibinfo{author}{\bibfnamefont{P.}~\bibnamefont{Danielewicz}} \bibnamefont{and}
  \bibinfo{author}{\bibfnamefont{G.}~\bibnamefont{Bertsch}},
  \bibinfo{journal}{Nucl. Phys. A} \textbf{\bibinfo{volume}{533}},
  \bibinfo{pages}{712} (\bibinfo{year}{1991}).

\bibitem[{\citenamefont{Mattiello et~al.}(1997)\citenamefont{Mattiello, Sorge,
  Stoecker, and Greiner}}]{Mattiello:1996gq}
\bibinfo{author}{\bibfnamefont{R.}~\bibnamefont{Mattiello}},
  \bibinfo{author}{\bibfnamefont{H.}~\bibnamefont{Sorge}},
  \bibinfo{author}{\bibfnamefont{H.}~\bibnamefont{Stoecker}}, \bibnamefont{and}
  \bibinfo{author}{\bibfnamefont{W.}~\bibnamefont{Greiner}},
  \bibinfo{journal}{Phys. Rev. C} \textbf{\bibinfo{volume}{55}},
  \bibinfo{pages}{1443} (\bibinfo{year}{1997}), \eprint{nucl-th/9607003}.

\bibitem[{\citenamefont{Blum et~al.}(2017)\citenamefont{Blum, Ng, Sato, and
  Takimoto}}]{Blum:2017qnn}
\bibinfo{author}{\bibfnamefont{K.}~\bibnamefont{Blum}},
  \bibinfo{author}{\bibfnamefont{K.~C.~Y.} \bibnamefont{Ng}},
  \bibinfo{author}{\bibfnamefont{R.}~\bibnamefont{Sato}}, \bibnamefont{and}
  \bibinfo{author}{\bibfnamefont{M.}~\bibnamefont{Takimoto}},
  \bibinfo{journal}{Phys. Rev.} \textbf{\bibinfo{volume}{D96}},
  \bibinfo{pages}{103021} (\bibinfo{year}{2017}), \eprint{1704.05431}.

\bibitem[{\citenamefont{Blum and Takimoto}(2019)}]{Blum:2019suo}
\bibinfo{author}{\bibfnamefont{K.}~\bibnamefont{Blum}} \bibnamefont{and}
  \bibinfo{author}{\bibfnamefont{M.}~\bibnamefont{Takimoto}},
  \bibinfo{journal}{Phys. Rev. C} \textbf{\bibinfo{volume}{99}},
  \bibinfo{pages}{044913} (\bibinfo{year}{2019}), \eprint{1901.07088}.

\bibitem[{\citenamefont{Acharya et~al.}(2018{\natexlab{a}})}]{Acharya:2017bso}
\bibinfo{author}{\bibfnamefont{S.}~\bibnamefont{Acharya}} \bibnamefont{et~al.}
  (\bibinfo{collaboration}{ALICE}), \bibinfo{journal}{Nucl. Phys. A}
  \textbf{\bibinfo{volume}{971}}, \bibinfo{pages}{1}
  (\bibinfo{year}{2018}{\natexlab{a}}), \eprint{1710.07531}.

\bibitem[{\citenamefont{Andronic et~al.}(2018)\citenamefont{Andronic,
  Braun-Munzinger, Redlich, and Stachel}}]{Andronic:2017pug}
\bibinfo{author}{\bibfnamefont{A.}~\bibnamefont{Andronic}},
  \bibinfo{author}{\bibfnamefont{P.}~\bibnamefont{Braun-Munzinger}},
  \bibinfo{author}{\bibfnamefont{K.}~\bibnamefont{Redlich}}, \bibnamefont{and}
  \bibinfo{author}{\bibfnamefont{J.}~\bibnamefont{Stachel}},
  \bibinfo{journal}{Nature} \textbf{\bibinfo{volume}{561}},
  \bibinfo{pages}{321} (\bibinfo{year}{2018}), \eprint{1710.09425}.

\bibitem[{\citenamefont{Vovchenko et~al.}(2018)\citenamefont{Vovchenko,
  Dönigus, and Stoecker}}]{Vovchenko:2018fiy}
\bibinfo{author}{\bibfnamefont{V.}~\bibnamefont{Vovchenko}},
  \bibinfo{author}{\bibfnamefont{B.}~\bibnamefont{Dönigus}}, \bibnamefont{and}
  \bibinfo{author}{\bibfnamefont{H.}~\bibnamefont{Stoecker}},
  \bibinfo{journal}{Phys. Lett. B} \textbf{\bibinfo{volume}{785}},
  \bibinfo{pages}{171} (\bibinfo{year}{2018}), \eprint{1808.05245}.

\bibitem[{\citenamefont{Bellini and Kalweit}(2019)}]{Bellini:2018epz}
\bibinfo{author}{\bibfnamefont{F.}~\bibnamefont{Bellini}} \bibnamefont{and}
  \bibinfo{author}{\bibfnamefont{A.~P.} \bibnamefont{Kalweit}},
  \bibinfo{journal}{Phys. Rev. C} \textbf{\bibinfo{volume}{99}},
  \bibinfo{pages}{054905} (\bibinfo{year}{2019}), \eprint{1807.05894}.

\bibitem[{\citenamefont{Chen et~al.}(2018)\citenamefont{Chen, Keane, Ma, Tang,
  and Xu}}]{Chen:2018tnh}
\bibinfo{author}{\bibfnamefont{J.}~\bibnamefont{Chen}},
  \bibinfo{author}{\bibfnamefont{D.}~\bibnamefont{Keane}},
  \bibinfo{author}{\bibfnamefont{Y.-G.} \bibnamefont{Ma}},
  \bibinfo{author}{\bibfnamefont{A.}~\bibnamefont{Tang}}, \bibnamefont{and}
  \bibinfo{author}{\bibfnamefont{Z.}~\bibnamefont{Xu}}, \bibinfo{journal}{Phys.
  Rept.} \textbf{\bibinfo{volume}{760}}, \bibinfo{pages}{1}
  (\bibinfo{year}{2018}), \eprint{1808.09619}.

\bibitem[{\citenamefont{Xu and Rapp}(2019)}]{Xu:2018jff}
\bibinfo{author}{\bibfnamefont{X.}~\bibnamefont{Xu}} \bibnamefont{and}
  \bibinfo{author}{\bibfnamefont{R.}~\bibnamefont{Rapp}},
  \bibinfo{journal}{Eur. Phys. J. A} \textbf{\bibinfo{volume}{55}},
  \bibinfo{pages}{68} (\bibinfo{year}{2019}), \eprint{1809.04024}.

\bibitem[{\citenamefont{Oliinychenko et~al.}(2019)\citenamefont{Oliinychenko,
  Pang, Elfner, and Koch}}]{Oliinychenko:2018ugs}
\bibinfo{author}{\bibfnamefont{D.}~\bibnamefont{Oliinychenko}},
  \bibinfo{author}{\bibfnamefont{L.-G.} \bibnamefont{Pang}},
  \bibinfo{author}{\bibfnamefont{H.}~\bibnamefont{Elfner}}, \bibnamefont{and}
  \bibinfo{author}{\bibfnamefont{V.}~\bibnamefont{Koch}},
  \bibinfo{journal}{Phys. Rev. C} \textbf{\bibinfo{volume}{99}},
  \bibinfo{pages}{044907} (\bibinfo{year}{2019}), \eprint{1809.03071}.

\bibitem[{\citenamefont{Bellini et~al.}(2021)\citenamefont{Bellini, Blum,
  Kalweit, and Puccio}}]{Bellini:2020cbj}
\bibinfo{author}{\bibfnamefont{F.}~\bibnamefont{Bellini}},
  \bibinfo{author}{\bibfnamefont{K.}~\bibnamefont{Blum}},
  \bibinfo{author}{\bibfnamefont{A.~P.} \bibnamefont{Kalweit}},
  \bibnamefont{and} \bibinfo{author}{\bibfnamefont{M.}~\bibnamefont{Puccio}},
  \bibinfo{journal}{Phys. Rev. C} \textbf{\bibinfo{volume}{103}},
  \bibinfo{pages}{014907} (\bibinfo{year}{2021}), \eprint{2007.01750}.

\bibitem[{\citenamefont{Mrowczynski}(1992)}]{Mrowczynski:1992gc}
\bibinfo{author}{\bibfnamefont{S.}~\bibnamefont{Mrowczynski}},
  \bibinfo{journal}{Phys. Lett. B} \textbf{\bibinfo{volume}{277}},
  \bibinfo{pages}{43} (\bibinfo{year}{1992}).

\bibitem[{\citenamefont{Lednicky et~al.}(1996)\citenamefont{Lednicky,
  Lyuboshits, Erazmus, and Nouais}}]{Lednicky:1995vk}
\bibinfo{author}{\bibfnamefont{R.}~\bibnamefont{Lednicky}},
  \bibinfo{author}{\bibfnamefont{V.~L.} \bibnamefont{Lyuboshits}},
  \bibinfo{author}{\bibfnamefont{B.}~\bibnamefont{Erazmus}}, \bibnamefont{and}
  \bibinfo{author}{\bibfnamefont{D.}~\bibnamefont{Nouais}},
  \bibinfo{journal}{Phys. Lett. B} \textbf{\bibinfo{volume}{373}},
  \bibinfo{pages}{30} (\bibinfo{year}{1996}).

\bibitem[{\citenamefont{Maj and Mrowczynski}(2009)}]{Maj:2009ue}
\bibinfo{author}{\bibfnamefont{R.}~\bibnamefont{Maj}} \bibnamefont{and}
  \bibinfo{author}{\bibfnamefont{S.}~\bibnamefont{Mrowczynski}},
  \bibinfo{journal}{Phys. Rev. C} \textbf{\bibinfo{volume}{80}},
  \bibinfo{pages}{034907} (\bibinfo{year}{2009}), \eprint{0903.0111}.

\bibitem[{\citenamefont{Nagle et~al.}(1994)\citenamefont{Nagle, Kumar, Bennett,
  Diebold, Pope, Sorge, and Sullivan}}]{Nagle:1994hm}
\bibinfo{author}{\bibfnamefont{J.~L.} \bibnamefont{Nagle}},
  \bibinfo{author}{\bibfnamefont{B.~S.} \bibnamefont{Kumar}},
  \bibinfo{author}{\bibfnamefont{M.~J.} \bibnamefont{Bennett}},
  \bibinfo{author}{\bibfnamefont{G.~E.} \bibnamefont{Diebold}},
  \bibinfo{author}{\bibfnamefont{J.~K.} \bibnamefont{Pope}},
  \bibinfo{author}{\bibfnamefont{H.}~\bibnamefont{Sorge}}, \bibnamefont{and}
  \bibinfo{author}{\bibfnamefont{J.~P.} \bibnamefont{Sullivan}},
  \bibinfo{journal}{Phys. Rev. Lett.} \textbf{\bibinfo{volume}{73}},
  \bibinfo{pages}{1219} (\bibinfo{year}{1994}).

\bibitem[{\citenamefont{Bierlich et~al.}(2022)}]{Bierlich:2022pfr}
\bibinfo{author}{\bibfnamefont{C.}~\bibnamefont{Bierlich}} \bibnamefont{et~al.}
  (\bibinfo{year}{2022}), \eprint{2203.11601}.

\bibitem[{\citenamefont{Ferreres-Solé and
  Sjöstrand}(2018)}]{Ferreres-Sole:2018vgo}
\bibinfo{author}{\bibfnamefont{S.}~\bibnamefont{Ferreres-Solé}}
  \bibnamefont{and}
  \bibinfo{author}{\bibfnamefont{T.}~\bibnamefont{Sjöstrand}},
  \bibinfo{journal}{Eur. Phys. J. C} \textbf{\bibinfo{volume}{78}},
  \bibinfo{pages}{983} (\bibinfo{year}{2018}), \eprint{1808.04619}.

\bibitem[{\citenamefont{Acharya et~al.}(2020{\natexlab{a}})}]{Acharya:2020dfb}
\bibinfo{author}{\bibfnamefont{S.}~\bibnamefont{Acharya}} \bibnamefont{et~al.}
  (\bibinfo{collaboration}{ALICE}), \bibinfo{journal}{Phys. Lett. B}
  \textbf{\bibinfo{volume}{811}}, \bibinfo{pages}{135849}
  (\bibinfo{year}{2020}{\natexlab{a}}), \eprint{2004.08018}.

\bibitem[{\citenamefont{Acharya et~al.}(2018{\natexlab{b}})}]{ALICE:2017xrp}
\bibinfo{author}{\bibfnamefont{S.}~\bibnamefont{Acharya}} \bibnamefont{et~al.}
  (\bibinfo{collaboration}{ALICE}), \bibinfo{journal}{Phys. Rev. C}
  \textbf{\bibinfo{volume}{97}}, \bibinfo{pages}{024615}
  (\bibinfo{year}{2018}{\natexlab{b}}), \eprint{1709.08522}.

\bibitem[{\citenamefont{{ALICE collaboration}}(2022)}]{ALICE:2022jmr}
\bibinfo{author}{\bibnamefont{{ALICE collaboration}}}
  (\bibinfo{collaboration}{ALICE}) (\bibinfo{year}{2022}), \eprint{2211.15204}.

\bibitem[{\citenamefont{Acharya et~al.}(2021)}]{ALICE:2020hjy}
\bibinfo{author}{\bibfnamefont{S.}~\bibnamefont{Acharya}} \bibnamefont{et~al.}
  (\bibinfo{collaboration}{ALICE}), \bibinfo{journal}{Phys. Lett. B}
  \textbf{\bibinfo{volume}{819}}, \bibinfo{pages}{136440}
  (\bibinfo{year}{2021}), \eprint{2011.05898}.

\bibitem[{\citenamefont{Lees et~al.}(2014)}]{BaBar:2014ssg}
\bibinfo{author}{\bibfnamefont{J.~P.} \bibnamefont{Lees}} \bibnamefont{et~al.}
  (\bibinfo{collaboration}{BaBar}), \bibinfo{journal}{Phys. Rev. D}
  \textbf{\bibinfo{volume}{89}}, \bibinfo{pages}{111102}
  (\bibinfo{year}{2014}), \eprint{1403.4409}.

\bibitem[{\citenamefont{Asner et~al.}(2007)}]{CLEO:2006zjy}
\bibinfo{author}{\bibfnamefont{D.~M.} \bibnamefont{Asner}} \bibnamefont{et~al.}
  (\bibinfo{collaboration}{CLEO}), \bibinfo{journal}{Phys. Rev. D}
  \textbf{\bibinfo{volume}{75}}, \bibinfo{pages}{012009}
  (\bibinfo{year}{2007}), \eprint{hep-ex/0612019}.

\bibitem[{\citenamefont{Lednicky}(2009)}]{Lednicky:2005tb}
\bibinfo{author}{\bibfnamefont{R.}~\bibnamefont{Lednicky}},
  \bibinfo{journal}{Phys. Part. Nucl.} \textbf{\bibinfo{volume}{40}},
  \bibinfo{pages}{307} (\bibinfo{year}{2009}), \eprint{nucl-th/0501065}.

\bibitem[{\citenamefont{Schiff}(1963)}]{Schiff63}
\bibinfo{author}{\bibfnamefont{L.}~\bibnamefont{Schiff}},
  \emph{\bibinfo{title}{{Quantum Mechanics}}}
  (\bibinfo{publisher}{{McGraw-Hill, New York}}, \bibinfo{year}{1963}).

\bibitem[{\citenamefont{Lisa et~al.}(2005)\citenamefont{Lisa, Pratt, Soltz, and
  Wiedemann}}]{Lisa:2005dd}
\bibinfo{author}{\bibfnamefont{M.~A.} \bibnamefont{Lisa}},
  \bibinfo{author}{\bibfnamefont{S.}~\bibnamefont{Pratt}},
  \bibinfo{author}{\bibfnamefont{R.}~\bibnamefont{Soltz}}, \bibnamefont{and}
  \bibinfo{author}{\bibfnamefont{U.}~\bibnamefont{Wiedemann}},
  \bibinfo{journal}{Ann. Rev. Nucl. Part. Sci.} \textbf{\bibinfo{volume}{55}},
  \bibinfo{pages}{357} (\bibinfo{year}{2005}), \eprint{nucl-ex/0505014}.

\bibitem[{\citenamefont{Sorge et~al.}(1989)\citenamefont{Sorge, Stoecker, and
  Greiner}}]{Sorge:1989vt}
\bibinfo{author}{\bibfnamefont{H.}~\bibnamefont{Sorge}},
  \bibinfo{author}{\bibfnamefont{H.}~\bibnamefont{Stoecker}}, \bibnamefont{and}
  \bibinfo{author}{\bibfnamefont{W.}~\bibnamefont{Greiner}},
  \bibinfo{journal}{Nucl. Phys. A} \textbf{\bibinfo{volume}{498}},
  \bibinfo{pages}{567C} (\bibinfo{year}{1989}).

\bibitem[{\citenamefont{Case}(2008)}]{doi:10.1119/1.2957889}
\bibinfo{author}{\bibfnamefont{W.~B.} \bibnamefont{Case}},
  \bibinfo{journal}{American Journal of Physics} \textbf{\bibinfo{volume}{76}},
  \bibinfo{pages}{937} (\bibinfo{year}{2008}),
  \eprint{https://doi.org/10.1119/1.2957889},
  \urlprefix\url{https://doi.org/10.1119/1.2957889}.

\bibitem[{\citenamefont{Zhaba}(2017)}]{Zhaba:2017syr}
\bibinfo{author}{\bibfnamefont{V.~I.} \bibnamefont{Zhaba}}
  (\bibinfo{year}{2017}), \eprint{1706.08306}.

\bibitem[{\citenamefont{Hudson}(1974)}]{HUDSON1974249}
\bibinfo{author}{\bibfnamefont{R.}~\bibnamefont{Hudson}},
  \bibinfo{journal}{Reports on Mathematical Physics}
  \textbf{\bibinfo{volume}{6}}, \bibinfo{pages}{249} (\bibinfo{year}{1974}),
  ISSN \bibinfo{issn}{0034-4877},
  \urlprefix\url{https://www.sciencedirect.com/science/article/pii/003448777490007X}.

\bibitem[{\citenamefont{Soto and Claverie}(1983)}]{doi:10.1063/1.525607}
\bibinfo{author}{\bibfnamefont{F.}~\bibnamefont{Soto}} \bibnamefont{and}
  \bibinfo{author}{\bibfnamefont{P.}~\bibnamefont{Claverie}},
  \bibinfo{journal}{Journal of Mathematical Physics}
  \textbf{\bibinfo{volume}{24}}, \bibinfo{pages}{97} (\bibinfo{year}{1983}),
  \eprint{https://doi.org/10.1063/1.525607},
  \urlprefix\url{https://doi.org/10.1063/1.525607}.

\bibitem[{\citenamefont{Werner et~al.}(2006)\citenamefont{Werner, Liu, and
  Pierog}}]{Werner:2005jf}
\bibinfo{author}{\bibfnamefont{K.}~\bibnamefont{Werner}},
  \bibinfo{author}{\bibfnamefont{F.-M.} \bibnamefont{Liu}}, \bibnamefont{and}
  \bibinfo{author}{\bibfnamefont{T.}~\bibnamefont{Pierog}},
  \bibinfo{journal}{Phys. Rev. C} \textbf{\bibinfo{volume}{74}},
  \bibinfo{pages}{044902} (\bibinfo{year}{2006}), \eprint{hep-ph/0506232}.

\bibitem[{\citenamefont{Pierog et~al.}(2015)\citenamefont{Pierog, Karpenko,
  Katzy, Yatsenko, and Werner}}]{Pierog:2013ria}
\bibinfo{author}{\bibfnamefont{T.}~\bibnamefont{Pierog}},
  \bibinfo{author}{\bibfnamefont{I.}~\bibnamefont{Karpenko}},
  \bibinfo{author}{\bibfnamefont{J.~M.} \bibnamefont{Katzy}},
  \bibinfo{author}{\bibfnamefont{E.}~\bibnamefont{Yatsenko}}, \bibnamefont{and}
  \bibinfo{author}{\bibfnamefont{K.}~\bibnamefont{Werner}},
  \bibinfo{journal}{Phys. Rev. C} \textbf{\bibinfo{volume}{92}},
  \bibinfo{pages}{034906} (\bibinfo{year}{2015}), \eprint{1306.0121}.

\bibitem[{\citenamefont{Ostapchenko}(2011)}]{Ostapchenko:2010vb}
\bibinfo{author}{\bibfnamefont{S.}~\bibnamefont{Ostapchenko}},
  \bibinfo{journal}{Phys. Rev.} \textbf{\bibinfo{volume}{D83}},
  \bibinfo{pages}{014018} (\bibinfo{year}{2011}), \eprint{1010.1869}.

\bibitem[{\citenamefont{Ostapchenko}(2013)}]{Ostapchenko:2013pia}
\bibinfo{author}{\bibfnamefont{S.}~\bibnamefont{Ostapchenko}},
  \bibinfo{journal}{EPJ Web Conf.} \textbf{\bibinfo{volume}{52}},
  \bibinfo{pages}{02001} (\bibinfo{year}{2013}).

\bibitem[{\citenamefont{Acharya et~al.}(2020{\natexlab{b}})}]{Acharya:2020sfy}
\bibinfo{author}{\bibfnamefont{S.}~\bibnamefont{Acharya}} \bibnamefont{et~al.}
  (\bibinfo{collaboration}{ALICE}), \bibinfo{journal}{Eur. Phys. J. C}
  \textbf{\bibinfo{volume}{80}}, \bibinfo{pages}{889}
  (\bibinfo{year}{2020}{\natexlab{b}}), \eprint{2003.03184}.

\bibitem[{\citenamefont{Marietti et~al.}(2022)\citenamefont{Marietti, Pilloni,
  and Tamponi}}]{Marietti:2022jil}
\bibinfo{author}{\bibfnamefont{D.}~\bibnamefont{Marietti}},
  \bibinfo{author}{\bibfnamefont{A.}~\bibnamefont{Pilloni}}, \bibnamefont{and}
  \bibinfo{author}{\bibfnamefont{U.}~\bibnamefont{Tamponi}}
  (\bibinfo{year}{2022}), \eprint{2208.14185}.

\bibitem[{\citenamefont{Workman et~al.}(2022)}]{ParticleDataGroup:2022pth}
\bibinfo{author}{\bibfnamefont{R.~L.} \bibnamefont{Workman}}
  \bibnamefont{et~al.} (\bibinfo{collaboration}{Particle Data Group}),
  \bibinfo{journal}{PTEP} \textbf{\bibinfo{volume}{2022}},
  \bibinfo{pages}{083C01} (\bibinfo{year}{2022}).

\bibitem[{\citenamefont{Horst et~al.}(2023)\citenamefont{Horst, Barioglio,
  Bellini, Fabbietti, Pinto, Singh, and Tripathy}}]{Horst:2023oti}
\bibinfo{author}{\bibfnamefont{M.}~\bibnamefont{Horst}},
  \bibinfo{author}{\bibfnamefont{L.}~\bibnamefont{Barioglio}},
  \bibinfo{author}{\bibfnamefont{F.}~\bibnamefont{Bellini}},
  \bibinfo{author}{\bibfnamefont{L.}~\bibnamefont{Fabbietti}},
  \bibinfo{author}{\bibfnamefont{C.}~\bibnamefont{Pinto}},
  \bibinfo{author}{\bibfnamefont{B.}~\bibnamefont{Singh}}, \bibnamefont{and}
  \bibinfo{author}{\bibfnamefont{S.}~\bibnamefont{Tripathy}}
  (\bibinfo{year}{2023}), \eprint{2302.12696}.

\end{thebibliography}

\end{document}